\documentclass[lineno]{jfm_clean}

\usepackage{graphicx}
\usepackage{newtxtext}
\usepackage{newtxmath}
\usepackage{natbib}
\usepackage{hyperref}
\usepackage{mathtools}

\hypersetup{
	colorlinks,
	citecolor=blue,      
	filecolor=red,
	linkcolor=blue,
	urlcolor=blue,
	hyperfigures
}

\newcommand{\RomanNumeralCaps}[1]
\linenumbers
\newcommand{\mbf}{\boldsymbol}
\newcommand{\nb}{\bnabla}
\newcommand{\nbp}{\bnabla_\parallel}
\newcommand{\dd}{\mathrm{d}}
\newcommand{\ii}{\mathrm{i}}
\newcommand{\ee}{\mathrm{e}}

\newcommand{\pdif}[2]{\dfrac{\partial #1}{\partial #2}}

\newcommand{\etheta}{\hat{\mbf{e}}_\theta}
\newcommand{\ephi}{\hat{\mbf{e}}_\phi}

\newcommand{\me}{\hat{\mbf{e}}}

\newcommand{\mm}{\mbf{m}}
\newcommand{\bmm}{\bar{\mbf{m}}}

\definecolor{Unsure}{RGB}{10, 70, 200}

\title{Collective dynamics of active suspensions on curved viscous interfaces}

\author{Yuzhu Chen\aff{1},
  Vishal P. Patil\aff{2}
 \and David Saintillan\aff{1}\corresp{\email{dstn@ucsd.edu}}}

\affiliation{\aff{1}Department of Mechanical and Aerospace Engineering, University of California San Diego, La Jolla, CA 92093, USA 
\aff{2}Department of Mathematics, University of California San Diego, La Jolla, CA 92093, USA}

\begin{document}
\maketitle

\begin{abstract}
Self-propelled particles can navigate complex environments, including viscous fluid interfaces with curved geometries. 
In this work, we study the emergent dynamics of a suspension of self-propelled particles confined to a stationary curved viscous interface. 
The evolution of the particle configurations is modeled using the Fokker-Planck equation on the curved surface, formulated using Cartan's moving frame method, and coupled to the bulk and surface Stokes equations with flows driven by an interfacial nematic active stress. 
Specifically, for a spherical vesicle, the flow field and the distribution of the particles are analyzed theoretically and numerically within the framework of spin-weighted functions and spin-weighted spherical harmonics, which provide a natural geometric description of the probability distribution function on the sphere. 
A linear stability analysis about the uniform, isotropic state is performed and predicts a finite-wavelength instability, with mode selection arising from the competition between the vesicle radius and the Saffman-Delbr\"uck length. 
This instability and the associated mode-selection mechanism are also confirmed in nonlinear numerical simulations using a pseudo-spectral method based on spin-weighted spherical harmonics.

\end{abstract}

\begin{keywords}
\end{keywords}


\section{Introduction}
\label{sec:headings}
Active suspensions of self-propelled particles have been extensively studied both experimentally and theoretically over the past few decades due to their rich phenomenology, including anomalous rheology \citep{saintillan2018rheology} and spontaneous collective dynamics \citep{koch2011collective}, also known as bacterial turbulence \citep{aditi2002hydrodynamic, sokolov2012physical, wensink2012meso,dunkel2013fluid}. 
The onset of bacterial turbulence can be explained by a mean-field kinetic theory \citep{saintillan2008instabilities}, in which rod-like active particles interact hydrodynamically through the far-field flow generated by the force dipoles they exert on the fluid as they propel themselves. 
In systems of extensile particles or so-called pushers, a linear instability occurs above a critical particle density, and the system transitions to a chaotic regime, characterized by large-scale nonlinear patterns involving fluctuations in both particle concentration and alignment. 
This linear instability in the 3D bulk case occurs at long wavelengths, with the most unstable length scale set by the system size. 
When active suspensions are confined to fluid interfaces, finite-wavelength instabilities have instead been predicted \citep{vskultety2024hydrodynamic,mahapatra2025self} and observed in experiments \citep{yin2026bacterial}, as the presence of multiple length scales modifies the instability and leads to mode selection. 

The transport of rod-like particles on curved viscous membranes is relevant to a range of cellular processes from cell division \citep{salbreux2009hydrodynamics} to the aggregation of BAR proteins \citep{simunovic2013linear}. 
Although several existing works have phenomenonologically modeled a continuum of rod-like particles as two-dimensional active polar and nematic fluids in curved spaces \citep{salbreux2022theory,al2023morphodynamics}, a first-principles hydrodynamic theory for a thin layer of rod-like particles confined to a curved viscous membrane embedded in surrounding fluids remains absent. 
Historically, the mobility of a single particle in a viscous membrane surrounded by 3D bulk fluids was first studied theoretically by Saffman and Delbr\"uck \citep{saffman1975brownian,saffman1976brownian}, who introduced the Saffman-Delbr\"uck length as the ratio between the 2D membrane viscosity and the 3D bulk viscosity so as to resolve the Stokes paradox. 
Subsequently, the mobility of rod-like inclusions and flexible filaments on flat membranes were studied \citep{levine2004mobility,levine2004dynamics}, and the hydrodynamics of a single filament embedded in a membrane with spherical geometry was explored \citep{shi2022hydrodynamics}. 
At the continuum level, the kinetic theory of bulk suspensions of \cite{saintillan2008instabilities} was recently extended to flat viscous membranes surrounded by 3D viscous fluids \citep{mahapatra2025self}. An analogous theory for suspensions confined to a curved viscous interface has yet to be developed. 

In this work, we address this gap and extend the kinetic theory for bulk dilute active suspensions of \cite{saintillan2008instabilities} to stationary curved viscous interfaces. 
We start from the coupled bulk and surface Stokes equations, and model the active stress exerted by the rod-like particles as the ensemble average of a distribution of force dipoles on the interface. 
The probability distribution function of the particle configurations is governed by the Fokker-Planck equation on the curved interface, which is formulated using Cartan's moving frames to separate the contribution of tangential transport from the rotational dynamics. 
We focus specifically on the case of spherical membranes and highlight the role of the dimensionless parameters defined by the ratios of the membrane radius to the Saffman-Delbr\"uck lengths.
We further elucidate their role in mode selection and the overall dynamics of the coupled system by means of a linear stability analysis and nonlinear numerical simulations. 

One major difficulty in both the theoretical analysis and numerical solution of this system arises from the fact that the Fokker-Planck equation is defined on the unit tangent bundle of the sphere \citep{lee2009manifolds}, which cannot be covered using a single coordinate chart. 
Furthermore, coordinate singularities at the poles make it difficult to apply (pseudo-)spectral methods with standard spherical harmonics, as these can lead to spurious oscillations and loss of convergence. 
To overcome these difficulties, we adopt the framework of spin-weighted functions and spin-weighted spherical harmonics (SWSHs), which was first introduced by \cite{newman1966note} for the study of the asymptotic behavior of gravitational fields, and was later applied in various areas such as cosmology \citep{thorne1980multipole}, geoscience \citep{michel2020mathematical}, computer graphics \citep{yi2024spin}, and active fluids \citep{mickelin2018anomalous,supekar2020linearly}. 
We show that the Fourier components of the probability density function with respect to the orientation angle can be interpreted as spin-weighted functions, and therefore can be expanded using spin-weighted spherical harmonics, which form a complete, orthornormal basis for $L^2$ spin-weighted functions. 
Such a description provides a natural framework to analyze a distribution of rod-like particles on curved surfaces, and subsequently leads to a pseudo-spectral method for solving the coupled systems numerically. 

The remainder of this paper is structured as follows. 
We first present the theoretical model in \S \ref{sec:problem def}, where we formulate the coupled Stokes and Fokker-Planck equations on a general curved viscous interface before specializing it to the spherical case. 
We also derive the orientational moment hierarchy equations in terms of Fourier coefficients and obtain a linear relation between the SWSH coefficients of the velocity field and probability density function. 
We then analyze the linear stability analysis of this system about the uniform isotropic state in \S \ref{sec:LSA}. A pseudo-spectral numerical method is introduced in \S \ref{sec:num method} to solve the full nonlinear system.  
Results from numerical simulations are then discussed in \S \ref{sec:num results}, where the dynamics of the concentration, polarity, and nematic order parameter fields are analyzed in light of the linear stability predictions. 
We further investigate the energy spectra of these fields, as well as the total system entropy production and its constituents.  
We conclude and discuss potential extensions in \S \ref{sec:conclusion}.

\section{Problem definition} \label{sec:problem def}
\subsection{Governing equations}
\begin{figure}
\centerline{\includegraphics[width=\textwidth]{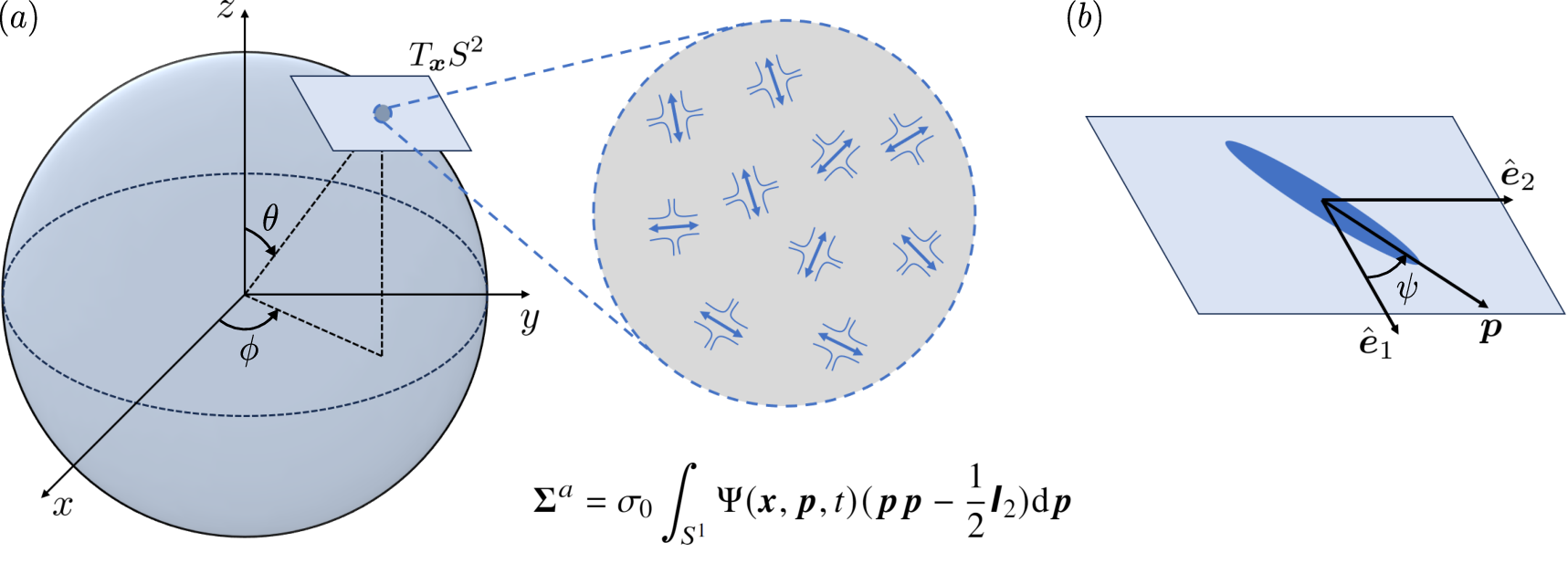}}
  \caption{($a$) Geometry of the system: an active suspension of rod-like particles is confined to a viscous interface (shown here as a sphere) separating two viscous fluids, where it exerts an active surface stress $\mbf{\Sigma}^a$. ($b$) In the tangent plane, the orientation $\boldsymbol{p}$ of a particle can be parametrized by the angle $\psi$, with respect to a frame $\{\me_1, \me_2\}$.}
\label{fig:figure 1}
\end{figure}

We consider a mean-field description of a thin layer of microswimmers, restricted to a curved stationary viscous incompressible membrane of 2D viscosity $\eta$, surrounded by a 3D bulk fluid of viscosity $\mu^{\text{in}}$ ($\mu^{\text{out}}$) inside (outside), as sketched in figure \ref{fig:figure 1} in the case of a spherical surface. 

The fluid velocity $\mbf{u}^{\text{in,out}}$ and pressure $p^{\text{in,out}}$ in the bulk domains satisfy the Stokes equations:
\begin{equation}
  - \mu^{\text{in,out}} \nabla^2\mbf{u}^{\text{in,out}} + \bnabla p^{\text{in,out}} = \boldsymbol{0},
  \quad \bnabla \bcdot \mbf{u}^{\text{in,out}} = 0, \quad \text{for } \mbf{x}\in V^{\text{in,out}}. 
\end{equation}
For the membrane, we start our formulation with a general 2D Riemannian manifold $(\mathcal{M},g)$, with coordinate basis $\mbf{g}^\alpha$ and Levi-Civita connection $\nabla_\alpha$, as discussed in detail by \cite{henle2010hydrodynamics} and \cite{samanta2021vortex}. The momentum and mass conservation equations on the membrane read
\begin{eqnarray}
     - \eta(\Delta_{C} + K)\mbf{u} + \nbp \gamma = \llbracket\mbf{T}\rrbracket+\mbf{f}^{a}  ,  \quad  \nbp \bcdot \mbf{u} = 0, \quad \text{for } \mbf{x}\in \mathcal{M}, 
\end{eqnarray}
where $\mbf{u}$ is the surface velocity, $\Delta_{C}=\nabla_\alpha \nabla^\alpha$ is the Connection Laplacian (or Bochner Laplacian with a minus sign) on the surface, $K$ is the Gaussian curvature, $\nbp = \mbf{g}^\alpha \nabla_\alpha$ is the covariant derivative on the surface, and $\gamma$ is the local surface pressure. 
We assume that the no-slip boundary condition
\begin{align}
    \mbf{u} = \mbf{u}^{\text{in}}|_{\mbf{x}\in\mathcal{M}} = \mbf{u}^{\text{out}}|_{\mbf{x}\in\mathcal{M}}
\end{align}
is satisfied at the membrane \citep{arroyo2009relaxation}. 
Viscous stresses in the bulk fluids result in the traction jump
\begin{eqnarray}
    \llbracket\mbf{T}\rrbracket = \mbf{N}\bcdot \llbracket -p\mathsfbi{I}_{\,3} + \mu(\nb\mbf{u} + \nb \mbf{u}^\top) \rrbracket,
\end{eqnarray}
where we have introduced the notation $\llbracket f\rrbracket = f^{\text{out}} - f^{\text{in}}$, $\boldsymbol{N}$ is a unit normal vector pointing from the inside towards the outside, and $\mathsfbi{I}_{\,n}$ is the identity matrix in $\mathbb{R}^n$. Finally, the microswimmers on the surface exert a force density  $\mbf{f}^{a} = \nbp \bcdot \mbf{\Sigma}^{a}$ resulting from the active surface stress $\mbf{\Sigma}^{a}$, whose form we discuss further below. Other notation and background on surface differential geometry are provided in Appendix \ref{appA}. 

The distribution of the microswimmers on the membrane is captured by the probability density function $\Psi(\mbf{x},\mbf{p},t)$ of finding a particle at position $\mbf{x} \in \mathcal{M}$ with orientation $\mbf{p}\in T_{\mbf{x}} \mathcal{M}$ where $|\mbf{p}|=1$. 
Thus, $\Psi(\mbf{x},\mbf{p},t)$ is a scalar function defined on the surface's unit tangent bundle $T^1 \mathcal{M}$ \citep{lee2009manifolds}, whose elements are pairs $(\mbf{x},\mbf{p}) \in T^1 \mathcal{M}$. 
After introducing a set of orthonormal frames $\{\me_1, \me_2\}\in T_{\mbf{x}} \mathcal{M}$, the director $\mbf{p}$ can be parameterized as $\mbf{p} = \cos\psi \me_1 + \sin\psi \me_2$ with $\psi\in[0,2\upi)$. 
Analogous to the flat bulk case \citep{saintillan2008instabilities}, the time evolution of the microswimmers is governed by the Fokker-Planck equation on the tangent bundle:
\begin{eqnarray}
    \pdif{\Psi}{t} = -\nbp \bcdot(\dot{\mbf{x}}_d\Psi) + D_t \Delta_{LB}\Psi  -\partial_{\psi}(\dot{\psi}_d\Psi) + D_r \partial^2_{\psi} \Psi, 
\end{eqnarray}
where $\Delta_{LB}$ denotes the Laplace-Beltrami operator.
The deterministic fluxes are given by
\begin{align}
    & \dot{\mbf{x}}_d = v_0 \mbf{p} + \mbf{u}, \\
    & \dot{\psi}_d = - \omega_{12}(\dot{\mbf{x}}_d) + \mbf{p}^\perp \bcdot (\beta\mathsfbi{E} - \mathsfbi{W}) \bcdot \mbf{p}, 
\end{align}
where $\mbf{p}^\perp = -\sin\psi \me_1 + \cos\psi \me_2$ is the counterclockwise $90^\circ$ rotation of $\mbf{p}$ in the tangent plane. 
Here $v_0$ is the self-propulsion speed, $D_t$ and $D_r$ are the surface translational and rotational diffusion coefficients, $\mathsfbi{E} = (\nbp \mbf{u} + \nbp \mbf{u}^\top)/2 $ is the surface strain-rate tensor, and $\mathsfbi{W} = (\nbp \mbf{u} - \nbp \mbf{u}^\top)/2$ is the surface vorticity tensor. 
The parameter $\beta$ is the flow-alignment parameter in Jeffery's equation that characterizes the shape of the particle, with $\beta=1$ for slender particles and $\beta=0$ for spherical particles. 
Different from the flat case, a connection term
\begin{align}
    \partial_\psi [- \omega_{12}(\dot{\mbf{x}}_d) \Psi]
\end{align}
appears and accounts for a rotational flux in $\psi$ due to the rotation of the frames. 
The same term also appears in the Fokker-Planck equation (38) for active Brownian particles without fluid interactions in \cite{castro2018active}. 
The derivations of these terms are contained in Appendix \ref{appA}, and one may refer to the books by \cite{do2012differential} and \cite{nakahara2018geometry} for a more detailed introduction to Cartan's moving frame method and the spin connection. 
The distribution function is normalized as
\begin{equation}
    \frac{1}{A} \int_{\mathcal{M}} \dd \mbf{x} \int_{S^1}  \dd \mbf{p} \Psi(\mbf{x},\mbf{p}, t) = n_p,
\end{equation}
where $A$ is the area of $\mathcal{M}$, $S^1$ is the unit circle, and $n_p=N_p/A$ is the mean surface number density of microswimmers. 

Given the distribution function and the stresslet strength $\sigma_0$, the active stress is obtained as
\begin{align}
    \mbf{\Sigma}^{a} & = \sigma_0 \int_{S^1} \Psi(\mbf{x},\mbf{p},t)(\mbf{p}\mbf{p} - \frac{1}{2}\mathsfbi{I}_{\,2}) \dd \mbf{p},
\end{align}
which can be viewed as the ensemble average in orientation of the stresslets exerted by the particles on the membrane \citep{saintillan2018rheology}. 
Thus the system is closed such that the active stress drives a fluid flow in the viscous membrane coupling to the bulk fluids, while the flow in turn transports and reorients the microswimmers through the flux terms in the Fokker-Planck equation.

\subsection{Non-dimensionalization}
There are multiple characteristic length scales in this system: length scales given by geometry, and the Saffman-Delbr\"uck lengths $L_{SD}^{\text{in,out}} = \eta / \mu^{\text{in,out}}$, which are ratios of the 2D membrane viscosity and the 3D bulk viscosities, and set the length scales for momentum transfer between the membrane and the bulk fluids. 
For membranes with general shapes, we note that not only the size of the membrane but also the local principal curvatures $\kappa_1$ and $\kappa_2$ may introduce multiple length scales. 
For simplicity, we specialize the remainder of the paper to a spherical membrane with radius $R$, so that the effect of curvature is captured by $1/R$ alone. 

Considering $N$ microswimmers on the sphere and choosing the characteristic time scale as $1/D_r$, the equations can be non-dimensionalized according to
\begin{eqnarray}
    \mbf{x}^\ast = \frac{\mbf{x}}{R}, \quad
    t^\ast = t D_r, \quad 
    \mbf{u}^\ast = \frac{\mbf{u}}{D_r R}, \quad 
    \Psi^\ast = \frac{\Psi}{N_p / 4\pi R^2},
\end{eqnarray}
with dimensionless parameters
\begin{eqnarray}
   &\displaystyle \alpha = \frac{N_p \sigma_0 }{4\pi R^2 \eta D_r}, \quad  U = \frac{v_0}{D_r R}, \quad D = \frac{D_t}{D_r R^2}, \ \\
   &\displaystyle\lambda^{\text{in}} = \frac{R}{\eta / \mu^{\text{in}}}, \quad
   \lambda^{\text{out}} = \frac{R}{\eta / \mu^{\text{out}}}. 
\end{eqnarray}
The activity of the system is characterized by the dimensionless activity $\alpha$ and the dimensionless self-propulsion speed $U$. 
The two dimensionless parameters $\lambda^{\text{in,out}}$ compare the scale of the system with the Saffman-Delbr\"uck lengths and play an important role in particle mobility and flow field topology, as shown in several previous works on inclusions in viscous membranes \citep{henle2008effect,henle2010hydrodynamics,samanta2021vortex,shi2022hydrodynamics,vona2025rotational}.
Specifically, the presence of bulk fluids can suppress the propagation of long-wavelength modes. 
In the limit of large Saffman-Delb\"uck length (small $\lambda^{\text{in,out}}$), the membrane is much more viscous than the bulk fluids, so that traction forces from the bulk are negligible and the system degenerates to the 2D case. 
As the membrane viscosity decreases, bulk fluids play a more important role in momentum transport. 
We will show that this effect will lead to mode selection in linear instabilities as well as nonlinear dynamics in the following sections.

With the scalings introduced above, the dimensionless governing equations (with stars dropped) can be summarized as follows:
\begin{eqnarray}
    &  \pdif{\Psi}{t} = -\nbp \bcdot(\dot{\mbf{x}}_d\Psi) + D \Delta_{LB}\Psi  -\partial_{\psi}(\dot{\psi}_d\Psi) + \partial^2_{\psi} \Psi, \quad \text{for } \mbf{x}\in \mathcal{M}, \label{eq:FPE} \\
    & \dot{\mbf{x}}_d = U \mbf{p} + \mbf{u}, \quad  \dot{\psi}_d = - \omega_{12}(\dot{\mbf{x}}_d) + \mbf{p}^\perp \bcdot (\beta\mathsfbi{E} - \mathsfbi{W}) \bcdot \mbf{p}, \\ [5pt]
    & - (\Delta_{C} + K)\mbf{u} + \nbp \gamma =\llbracket\mbf{T}\rrbracket+\nbp\bcdot\mbf{\Sigma}^{a}, \quad \nbp \bcdot \mbf{u} = 0, \quad \text{for } \mbf{x}\in \mathcal{M}, \\ [5pt]
    & - \nabla^2\mbf{u}^{\text{in,out}} + \bnabla p^{\text{in,out}} = \boldsymbol{0},
  \quad \bnabla \bcdot \mbf{u}^{\text{in,out}} = 0, \quad \text{for } \mbf{x}\in V^{\text{in,out}}. 
\end{eqnarray}

\subsection{Spin-weighted functions and solutions}
We use the standard spherical coordinates $(\theta,\phi) $ as the spatial coordinates on the sphere surface, with $\{\etheta,\ephi\}$ as the orthonormal basis. The Fokker-Planck equation \eqref{eq:FPE} can be written as
\begin{align}  \label{eq:FPE2}
    \pdif{\Psi}{t} = 
    & - \nbp \bcdot [(U\mbf{p} + \mbf{u})\Psi] 
    + D \Delta_{LB} \Psi + \pdif{^2\Psi}{\psi^2} \notag\\
    & + \pdif{}{\psi} [(U\cot\theta \sin\psi + \cos\theta u^\phi)\Psi - \mbf{p}^\perp \bcdot (\beta\mathsfbi{E} - \mathsfbi{W}) \bcdot \mbf{p} \Psi]. 
\end{align}
We note that in spherical coordinates, the spin connection terms contain geometric factors such as $\cot \theta$ that are singular at the poles, making it inefficient to expand them using the standard spherical harmonics. 
More fundamentally, with a single chart the local frame $\{\etheta,\ephi\}$ is not defined at the poles, preventing the orientation angle $\psi$ from being globally defined on $S^2$. 
However, the unit tangent bundle of $S^2$ can be identified with the rotation group $SO(3)$ \citep{marsden2013introduction,nestruev2003smooth}, so the Wigner-D matrix, or equivalently, the spin-weighted spherical harmonics \citep{goldberg1967spin,price2024differentiable}, can naturally serve as a complete orthonormal basis for $L^2$-functions on this bundle. 
We will show that the frame singularities are then encoded in the spin-weighted functions that are Fourier coefficients of a Fourier expansion in $\psi$, with the spin weight reflecting the $U(1)$ symmetry in the unit tangent bundle. 
Moreover, with the $\eth,\bar{\eth}$ operators defined below, the connection terms can be absorbed into the covariant derivatives of spin-weighted functions. This  yields a sparse spectral representation of the equations, which is convenient for linear stability analysis and numerical simulations. 

The spin-weighted functions on $S^2$ \citep{newman1966note,goldberg1967spin} are functions that transform as
\begin{align}
    f_s \to \ee^{\ii s \alpha} f_s 
\end{align}
under a clockwise rotation of the tangent planes by $\alpha$. It is convenient to write this as $\mm \to \ee^{\ii \alpha} \mm$, where $\mm = (\etheta + \ii \ephi)/\sqrt{2}$, $\smash{\bmm = (\etheta - \ii \ephi)/\sqrt{2}}$, is the null basis (see Appendix \ref{appB} for a more detailed introduction to spin-weighted spherical harmonics). 
The Fourier expansion of the probability density distribution with respect to $\psi$ is
\begin{align}
    \Psi(\theta,\phi,\psi,t) = \sum_n \Psi_n(\theta,\phi,t) \ee^{\ii n \psi}.
\end{align}
Since $\Psi$ itself should be a scalar function that does not depend on the choice of frames, after a clockwise rotation of frames $\psi \to \psi' = \psi + \alpha$,
\begin{align}
    \Psi'(\theta,\phi,\psi',t) = \sum_n \Psi_n'(\theta,\phi,t) \ee^{\ii n (\psi+\alpha)}
\end{align}
should remain unchanged. 
This requires $\Psi_n \to \Psi_n' = \ee^{-\ii n\alpha}\Psi_n$, so that the $n$-th Fourier coefficient $\Psi_n$ satisfies the definition of a spin-weighted function with spin weight $-n$, and thus can be expanded using the spin-$(-n)$ spherical harmonics. 

The orientational moments of the distribution function, by construction, are naturally diagonalized under the null basis. 
For instance, the concentration $c$, polarization $\mbf{P}$, and nematic tensor $\mathsfbi{Q}$ are expressed as
\begin{align}
    & c = \langle 1 \rangle = \int_0^{2\pi} \Psi(\theta,\phi,\psi,t) \dd \psi = 2\pi \Psi_0, \\
    & \mbf{P} = \frac{1}{c}\langle \mbf{p} \rangle = \frac{1}{c} \int_0^{2\pi} \Psi(\theta,\phi,\psi,t) \mbf{p} \dd \psi = \frac{\Psi_1 \mm + \Psi_{-1}\bmm}{\sqrt{2}\Psi_0}, \\ 
    & \mathsfbi{Q} = \frac{1}{c}\langle \mbf{p} \mbf{p} - \frac{1}{2}\mathsfbi{I}_{\,2} \rangle = \frac{1}{c}\int_0^{2\pi} \Psi(\theta,\phi,\psi,t) (\mbf{p} \mbf{p} - \frac{1}{2}\mathsfbi{I}_{\,2}) \dd \psi = \frac{\Psi_2 \mm\mm + \Psi_{-2}\bmm\bmm}{2\Psi_0}. 
\end{align}
Note that the quantity $\mm\mm$ ($\bmm\bmm$) has spin weight $2$ ($-2$). 
Combined with the coefficient $\Psi_2$ ($\Psi_{-2}$), the total spin weight of the nematic tensor is therefore zero, hence it is frame-invariant as required. 

The $\eth,\bar{\eth}$ operators are covariant differential operators acting on a spin-$s$ quantity $f$, defined by
\begin{align}
    & \eth f = - (\partial_\theta + \frac{\ii}{\sin\theta}\partial_\phi)f + s\cot \theta f, \\
    & \bar{\eth} f = - (\partial_\theta - \frac{\ii}{\sin\theta}\partial_\phi)f - s\cot \theta f.
\end{align}
The action of the $\eth$ operator raises the spin weight of $f$ by 1, while the action of the $\bar{\eth}$ operator lowers it by 1. 
With the operators and basis introduced above, the Fourier coefficients of the probability density function satisfy the moment hierarchy equations (see Appendix~\ref{sec:} for details of the derivation)
\begin{align}
    & \pdif{\Psi_n}{t} = \frac{U}{2}(\bar{\eth}\Psi_{n-1} + \eth \Psi_{n+1}) + \frac{D}{2 } (\eth\bar{\eth} + \bar{\eth}\eth) \Psi_n - n^2 \Psi_n \notag \\
    & \quad \quad \quad  + \frac{1}{\sqrt{2}} (u^+\bar{\eth}\Psi_n + u^- \eth \Psi_n) 
    + \frac{n}{2\sqrt{2}}
   [\beta(\eth u^+ \Psi_{n+2}
    - \bar{\eth} u^- \Psi_{n-2}) 
    - 2\eth u^- \Psi_n],  \label{eq:Psinequation}
\end{align}
where 
\begin{align}
    u^+ = \mbf{u}\bcdot \mm, \quad u^- = \mbf{u}\bcdot \bmm
\end{align}
are the projections of the velocity field to the null basis, and are quantities with spin weight $\pm 1$. 
At this stage, we can observe from equation~(\ref{eq:Psinequation}) that the self-propulsion term couples $\Psi_n$ with its closest neighboring moments  $\Psi_{n\pm 1}$, while the flow-alignment term couples it with the second-closest moments $\Psi_{n\pm 2}$. 

We can further expand the spin-weighted functions using the SWSHs as
\begin{align}
   & \Psi_n(\theta,\phi,t) = \sum_{l=|n|}^{\infty} \sum_{m=-l}^l a_{lm}^{(-n)}(t) {}_{-n}Y_{lm}(\theta,\phi), \\
   & u^\pm(\theta,\phi,t) = \sum_{l=1}^\infty \sum_{m=-l}^l u_{lm}^{\pm}(t) {}_{\pm 1}Y_{lm}(\theta,\phi),
\end{align}
with reality constraints
\begin{align}
    a_{lm}^{(-n)} = (-1)^{n-m} \bar{a}_{l,-m}^{(n)} \quad \text{and} \quad u_{lm}^{\pm} = (-1)^{1 -m} \bar{u}_{l,-m}^{\mp}. \label{eq:reality}
\end{align}
The surface incompressibility condition $\nbp\bcdot \mbf{u} = 0$ demands
\begin{align}
    u_{lm}^+ = u_{lm}^-. 
\end{align}
The velocity field in the bulk fluids can be obtained from Lamb's solution \citep{kim2013microhydrodynamics}. 
Upon applying the no-slip boundary condition and surface incompressibility, the traction jump is obtained as
\begin{align}
    \llbracket\mbf{T}\rrbracket = - \sum_{l=1}^\infty [\lambda^{\text{out}}(l+2) + \lambda^{\text{in}}(l-1)]\sum_{m=-l}^l u_{lm}^+ ({}_1 Y_{lm} \bar{\mbf{m}} + {}_{-1} Y_{lm} \mbf{m}).
\end{align}
The active force reads
\begin{align}
    \nbp \bcdot \mbf{\Sigma}^a = \frac{\pi\alpha}{\sqrt{2}}\sum_{l=2}^\infty \sqrt{(l+2)(l-1)}\sum_{m=-l}^{l}(a_{lm}^{(-2)}{}_{-1}Y_{lm} \mm - a_{lm}^{(2)}{}_{1}Y_{lm} \bmm),
\end{align}
and the viscous force on the membrane is
\begin{align}
    (\Delta_{C}+K)\mbf{u} = \sum_{l=1}^\infty
    [2-l(l+1)] \sum_{m=-l}^l u_{lm}^+({}_{1}Y_{lm} \bmm + {}_{-1}Y_{lm} \mm). 
\end{align}
See Appendix \ref{appC} for a more detailed derivation of these expressions. 
Applying the $\eth$ and $\bar{\eth}$ operators to the force balance on the membrane and eliminating the $\eth\bar{\eth}\gamma$ and $\bar{\eth}\eth \gamma$ terms, we can establish a linear relation between the $n=\pm 2$ coefficients of the probability density function and the velocity coefficients,
\begin{align}
    u_{lm}^{+} = u_{lm}^{-} = \frac{\pi\alpha \sqrt{(l-1)(l+2)}}{2\sqrt{2} s_l} (a_{lm}^{(2)} - a_{lm}^{(-2)}), \label{eq:uPDFcoeff}
\end{align}
where
\begin{align}
    s_l = [(l+2)\lambda^{\text{out}} + (l-1)\lambda^{\text{in}}] - 2+l(l+1)
\end{align}
is the same coefficient that appears in \cite{shi2022hydrodynamics} and \cite{samanta2021vortex}.

\section{Linear stability analysis} \label{sec:LSA}
\subsection{Linear stability around the uniform isotropic state}
We perform a linear stability analysis around the uniform, isotropic state ($\Psi=1/2\upi$) where the fluid velocity is zero ($\mbf{u}=\mbf{u}^{\mathrm{in}}=\mbf{u}^{\mathrm{out}}=\mathbf{0}$). 
Substituting the perturbed states
\begin{align}
    \Psi(\theta,\phi,\psi,t) = \frac{1}{2\pi} [1 + \delta\Psi(\theta,\phi,\psi,t)], \quad \mbf{u} = \delta\mbf{u} ,
\end{align}
into the Fokker-Planck equation (\ref{eq:FPE2}) yields the linearized equation 
\begin{align}
    \pdif{\delta\Psi}{t} = - \nbp \bcdot (U\mbf{p}\delta\Psi) + U\cot\theta \partial_\psi(\sin\psi \delta\Psi) + D \Delta_{LB} \delta \Psi + \partial_\psi^2 \delta\Psi + 2 \beta \mbf{p}\mbf{p}\boldsymbol{:}\delta \mathsfbi{E}. 
\end{align}
Since the SWSHs form a complete and orthonormal basis, we can use them as normal modes to expand the perturbations:
\begin{align}
    & \delta\Psi(\theta,\phi,\psi,t) = \sum_{n=-\infty}^{\infty} \sum_{l=|n|}^\infty \sum_{m=-l}^l \tilde{a}_{lm}^{(-n)} \; {}_{-n}Y_{lm}(\theta,\phi) \ee^{\ii n\psi + \sigma t}, \\
    & \delta u^{\pm}(\theta,\phi,t) = \sum_{l=1}^{\infty} \sum_{m=-l}^l \tilde{u}_{lm}^{\pm} \; {}_{\pm 1}Y_{lm}(\theta,\phi) \ee^{\sigma t},
\end{align}
and obtain the eigenvalue problem for the coefficients:
\begin{align}
\begin{split}
    \sigma \tilde{a}_{lm}^{(-n)} = &
    \frac{U}{2}[-\sqrt{(l-n+1)(l+n)}\tilde{a}_{lm}^{(-n+1)} + \sqrt{(l+n+1)(l-n)} \tilde{a}_{lm}^{(-n-1)} ] \\
    & - D[l(l+1)-n^2]\tilde{a}_{lm}^{(-n)} - n^2 \tilde{a}_{lm}^{(-n)} \\
    & - \frac{\beta\alpha}{8s_l}(l-1)(l+2)(\delta_{n,-2}-\delta_{n,2})(\tilde{a}_{lm}^{(n)} - \tilde{a}_{lm}^{(-n)}). 
    \end{split} \label{eq:LSA coeffs}
\end{align}
Note that there is no dependence on $m$, no coupling between different $l$'s, and the only coupling is between different spin weights $n$. 
This enables us to analyze the eigenspectra of the coefficients with different $l$'s independently. 
We also point out that, without the flow alignment term, the eigenvalue problem above is exactly the same as the SWSH spectral expansion for active Brownian particles with no interactions on a sphere \citep{castro2018active}, where the coefficient matrix before $\tilde{a}_{lm}^{(-n)}$ with fixed $l$ and $m$ is a tridiagonal matrix with negative eigenvalues. 
Therefore, the system can be linearly unstable only in the presence of the flow alignment term, which originates from the coupling between the nematic tensor and the surface strain-rate tensor, namely, the $n=\pm 2$ modes. 

We now analyze solutions of the eigenvalue problem \eqref{eq:LSA coeffs} for different fixed $l$'s. 
Note that the SWSHs are only nonzero when $l\geq |n|$, so for each value of $l$ the eigenvalue problem yields a $(2l+1)\times(2l+1)$ matrix for $|n|\leq l$. 
The physical meaning of this constraint can be interpreted as follows: 
A mode with spin weight $|n|$ cannot admit a spatial symmetry less than $l$, for example, there is no spatially homogeneous state ($l=0$) or state with $l=1$ on $S^2$ for a nematic field, of which the components are of spin weight $\pm 2$. 

\begin{figure}
\centerline{\includegraphics[width=\textwidth]{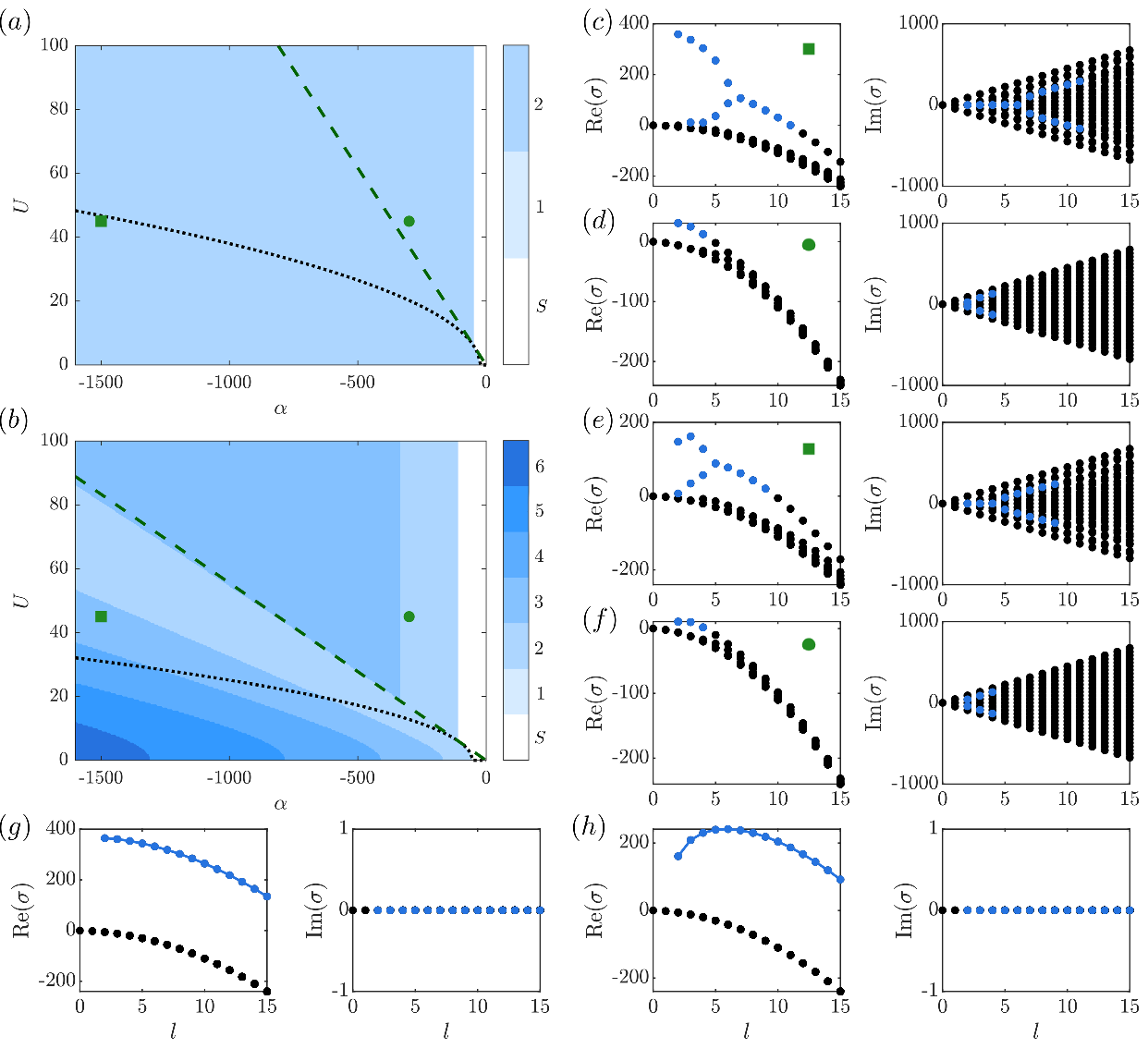}}
  \caption{($a,b$) Stability diagram in the $(U,\alpha)$ parameter space, where unstable regions are colored in blue according to the degree $l$ of the most unstable eigenmode, and stable regions ($S$) are shown in white. The black dotted line depicts the marginal condition for the first criterion in equation (\ref{eq:criterion}), while the dashed green line demarcates regions where the two dominant eigenvalues are real (below) versus complex conjugates (above). Parameter values are $D = 1$, 
  $\lambda=0.01$ in ($a$), and $D = 1$, 
  $\lambda=1$ in ($b$).  ($c$-$h$) Plots of the real and imaginary parts of the eigenvalues $\sigma$ as functions of degree $l$, where unstable eigenvalues are shown in blue, and stable eigenvalues in black. Parameter values are: 
  ($c$) $\alpha=-1500$, $U=45$, $D=1$, 
  $\lambda = 0.01$; 
  ($d$) $\alpha=-300$, $U=45$, $D=1$, 
  $\lambda = 0.01$; 
  ($e$) $\alpha=-1500$, $U=45$, $D=1$, 
  $\lambda = 1$; 
  ($f$) $\alpha=-300$, $U=45$, $D=1$, 
  $\lambda=1$;
  ($g$) $\alpha=-1500$, $U=0$, $D=1$, 
  $\lambda=0.01$;
  ($h$) $\alpha=-1500$, $U=0$, $D=1$, 
  $\lambda = 1$. Conditions for cases ($c$) through ($f$) are highlighted by the green symbols in panels ($a$) and ($b$). The unstable growth rates for cases ($g$) and ($h$) are analytical and given by equation (\ref{eq:shakersigma}). 
  Note that for each value of $l$ there are $2l+1$ eigenvalues. }
\label{fig:LSA}
\end{figure}

We start our discussion with $l=0$, for which the only choice of the spin weight is $n=0$ with azimuthal wavenumber $m=0$. The corresponding eigenvalue problem is
\begin{align}
    \sigma \tilde{a}_{00}^{(0)} = 0, 
\end{align}
which implies that a uniform fluctuation in concentration does not grow, as expected since the total number of particles should be conserved. 
Next, a perturbation with $l=1$ yields the eigenvalue problem
\begin{align}
    \sigma \begin{pmatrix}
        \tilde{a}_{1m}^{(-1)} \\[5pt]
        \tilde{a}_{1m}^{(0)} \\[5pt]
        \tilde{a}_{1m}^{(1)}
    \end{pmatrix}  
    = \begin{pmatrix}
        -D-1 & -\frac{\sqrt{2}}{2}U & 0 \\[5pt]
        \frac{\sqrt{2}}{2}U & -2D & -\frac{\sqrt{2}}{2}U \\[5pt]
        0 & \frac{\sqrt{2}}{2}U & -D-1
    \end{pmatrix}
    \begin{pmatrix}
        \tilde{a}_{1m}^{(-1)} \\[5pt]
        \tilde{a}_{1m}^{(0)} \\[5pt]
        \tilde{a}_{1m}^{(1)}
    \end{pmatrix} .
\end{align}
This yields the three eigenvalues 
\begin{align}
    \sigma_1 = -D-1, \quad \sigma_{2,3} = \frac{1}{2}[-(3D+1) \pm \sqrt{(3D+1)^2 - 8D(D+1) -4U^2}],
\end{align}
which all have negative real parts regardless of the choice of the parameters, implying that a perturbation with $l=1$ in either the concentration ($n=0$) or the polarity field ($n=\pm 1$) is also linearly stable. 
The first mode that can be unstable is $l=2$, where the eigenvalue problem becomes:
\begin{align}
    \sigma \begin{pmatrix}
        \tilde{a}_{2m}^{(-2)} \\[5pt]
        \tilde{a}_{2m}^{(-1)} \\[5pt]
        \tilde{a}_{2m}^{(0)} \\[5pt]
        \tilde{a}_{2m}^{(-1)} \\[5pt]
        \tilde{a}_{2m}^{(2)}
    \end{pmatrix}  
    = \begin{pmatrix}
        -2D-4-\frac{\beta\alpha}{2s_2} & -U & 0 & 0 & \frac{\beta\alpha}{2s_2} \\[5pt]
        U & -5D-1 & -\frac{\sqrt{6}}{2}U & 0 & 0 \\[5pt]
        0 & \frac{\sqrt{6}}{2}U & -6D & -\frac{\sqrt{6}}{2}U & 0 \\[5pt]
        0 & 0 & \frac{\sqrt{6}}{2}U & -5D-1 & -U \\[5pt]
        \frac{\beta\alpha}{2s_2} & 0 & 0 & U & -2D-4-\frac{\beta\alpha}{2s_2} 
    \end{pmatrix}
    \begin{pmatrix}
        \tilde{a}_{2m}^{(-2)} \\[5pt]
        \tilde{a}_{2m}^{(-1)} \\[5pt]
        \tilde{a}_{2m}^{(0)} \\[5pt]
        \tilde{a}_{2m}^{(-1)} \\[5pt]
        \tilde{a}_{2m}^{(2)}
    \end{pmatrix}, 
\end{align}
of which the characteristic polynomial can be rearranged as
\begin{align}
    \det \begin{pmatrix}
    -2D-4-\frac{\beta\alpha}{s_2}-\sigma & -U \\ U & -5D-1-\sigma
    \end{pmatrix}
    \det \begin{pmatrix}
    -6D-\sigma & -\sqrt{3}U & 0 \\
    \sqrt{3}U & -5D-1-\sigma & -U \\
    0 & U & -2D-4-\sigma
    \end{pmatrix} = 0
\end{align}
using an orthogonal transformation. 
Notice that the coefficient matrix can be decomposed into two blocks, with dimensions $2 \times 2$ and $3 \times 3$, respectively. 
From the Routh-Hurwitz criterion, the cubic part is always stable, so the condition for the $l=2$ mode to be unstable is only determined by the quadratic part, and reads
\begin{align}
    U^2 < -(5D+1)(2D+4+\frac{\beta\alpha}{s_2}) \quad {\text{or}}\quad 
    \frac{\beta\alpha}{s_2} < -7D-5. \label{eq:criterion}
\end{align}
For $l>2$, the unstable block is of dimension $l\times l$ and the eigenvalue problem becomes untractable analytically, so we calculate the dispersion relations numerically. 
We summarize the results in phase diagrams in the ($U,\alpha$) plane in 
figure \ref{fig:LSA}($a$) and ($b$), where we color regions according to the linear unstable modes, and the first criterion in equation (\ref{eq:criterion}) is shown as a dotted black line. 

A typical dispersion relation is plotted in figure \ref{fig:LSA}($c$) in the limit where $\lambda^{\text{in,out}}$ are small, corresponding to the situation where the membrane viscosity dominates and bulk viscous stresses are negligible. 
In this figure and in all the results that follow, we assume $\lambda^{\text{in}} = \lambda^{\text{out}}$ and denote the viscosity ratio simply by $\lambda$. 
The unstable branches (highlighted in blue) are similar to the bulk case \citep{saintillan2008instabilities}, where there is a long-wavelength instability with two positive real eigenvalues for small wavenumbers and two conjugate complex eigenvalues for large wavenumbers. 
As $\lambda$ increase, the maximum growth rate shifts to higher wavenumbers (figure \ref{fig:LSA}($e$)), resulting in mode selection in the linear regime, as also seen in the phase diagram of figure \ref{fig:LSA}($a$) and ($b$) with different values of $\lambda$. 
Moreover, we find that the maximum growth rate decreases as $\lambda$ increase and more viscous dissipation takes place in the bulk. 
We also note that the criterion $2U = -\beta\alpha/s_2 + 3D-3$, shown as a green dashed line in figure \ref{fig:LSA}($a$) and ($b$), determines the nature of unstable growth rates: 
Below this criterion (figure \ref{fig:LSA}($c$) and ($e$)), unstable low-wavenumber modes are characterized by two real positive growth rates; above the criterion, the growth rates become complex conjugate pairs with  positive real parts (figure \ref{fig:LSA}($d$) and ($f$)), suggesting the emergence of oscillatory solutions. 
In the degenerate case where $U=0$ (so-called ``shakers", \cite{ezhilan2013instabilities,stenhammar2017role,ohm2022weakly}), there is only one positive growth rate for each value of $l$ with the analytical expression (plotted as a solid line in figure \ref{fig:LSA}($g$) and ($h$)) 
\begin{align}
    \sigma_l = -[l(l+1)-4]D - 4 - \frac{\beta\alpha}{8s_l}(l-1)(l+2), \label{eq:shakersigma}
\end{align}
while the negative growth rates are given by $-[l(l+1)-n^2]D-n^2$ and capture the damping effect of translational and rotational diffusion.

\subsection{Eigenmodes}
The solution of the eigenvalue problem  \eqref{eq:LSA coeffs} also provides the eigenmodes of the distribution function, which can be used to reconstruct the polarity and  nematic tensor fields. 
The corresponding velocity field can also be reconstructed using equation \eqref{eq:uPDFcoeff}. 
As already mentioned, for each fixed $l$, the growth rates for all $m$'s are identical. 
However, the coefficients obtained from the eigenvector must satisfy the reality constraints \eqref{eq:reality}, which are not guaranteed after solving the eigenvalue problem. 
To satisfy the reality constraint, we use the linear combination of the eigenvectors $v_{m}^{(-n)}$ of the matrix given by each fixed $l$, and construct the eigenmodes as
\begin{align}
    \tilde{v}_{n,m} = c_m v_{m}^{(-n)} \;  + c_{-m} v_{-m}^{(n)}\;,
\end{align}
where the coefficients satisfy
\begin{align}
    c_m = (-1)^{m+1} \bar{c}_{-m}. 
\end{align}

\begin{figure}
\centerline{\includegraphics[width=\textwidth]{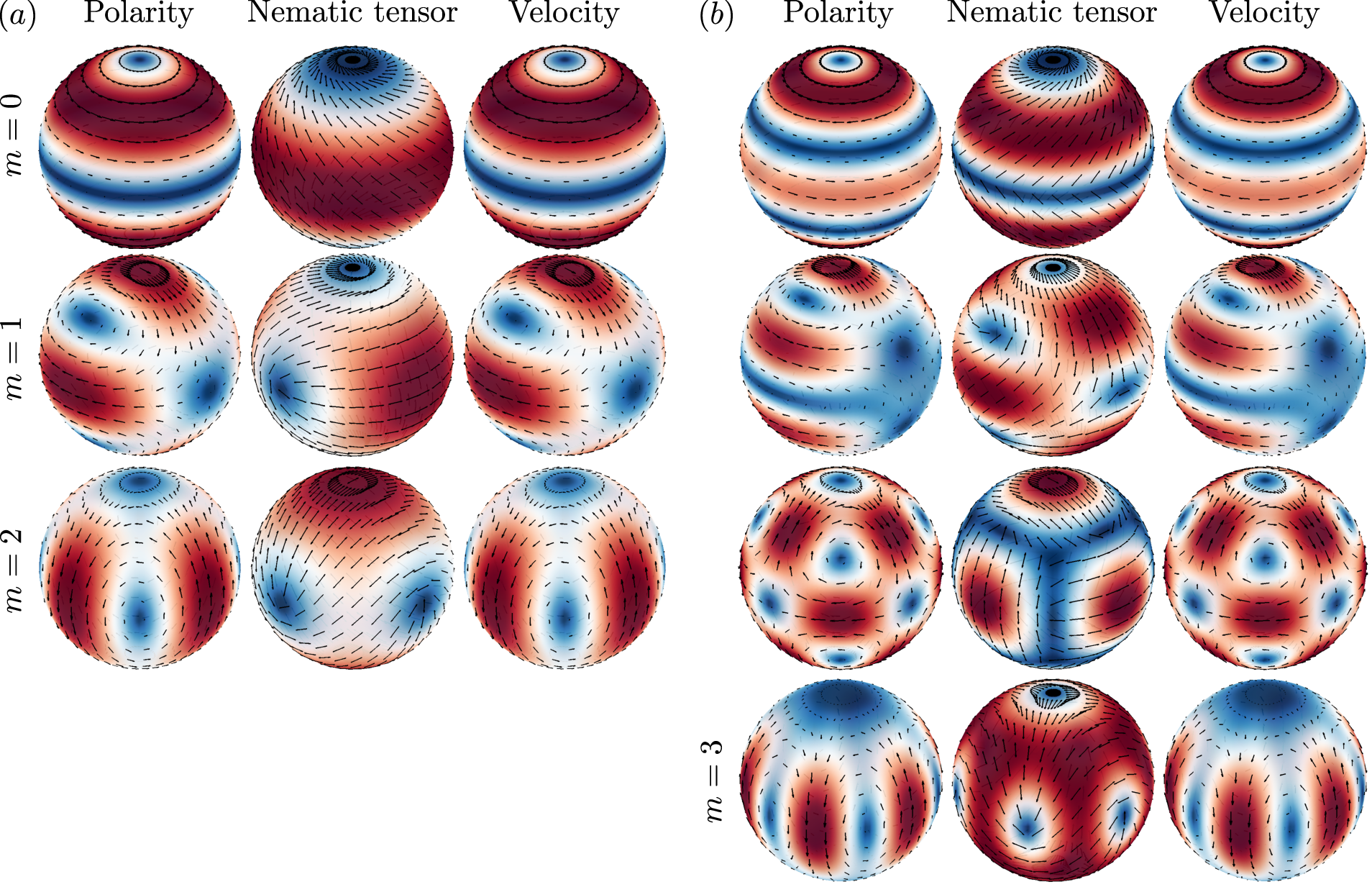}}
  \caption{Unstable eigenmodes for the polarity, nematic tensor and velocity fields, corresponding to degrees ($a$) $l=2$ and ($b$) $l=3$. Eigenmodes for distinct values of $m$ have identical growth rates. 
  Parameters values: ($a$) $\alpha=-1500$, $U=45$, $D=1$, 
  $\lambda=0.01$;
  ($b$) $\alpha=-1500$, $U=45$, $D=1$, 
  $\lambda=1$. 
  Note that the magnitudes of the eigenfunctions are defined up to an arbitrary scaling factor. }
\label{fig:eigenmodes}
\end{figure}

The unstable eigenmodes for $l=2$ are plotted in figure \ref{fig:eigenmodes}($a$), where we only show the $m\geq 0$ modes. 
The $m<0$ modes are the same as $-m$ for odd $m$ and are rotated by $\pi/n$ for even $m$. 
We can see from the eigenvalue problem \eqref{eq:LSA coeffs} that the relative magnitude of the polarity and the nematic tensor in the linear regime is determined by the ratio of $-[l(l+1)-4]D -4 - (l-1)(l+2)\beta\alpha/4s_l - \sigma_{\text{max}}$ to $U$. 
For $m=0$, the rotational symmetry yields an azimuthal shear flow in both the polarity and velocity fields, and the nematic tensor aligns with the principal axis of the surface strain-rate tensor under the shear flow, displaying two $+1$ defects at the poles. 
When $m=1$ and $2$, the nematic tensor instead exhibits four $+1/2$ defects equidistantly distributed along a great circle, with alternating directions, forming a baseball-like structure similar to that previously reported by \cite{vitelli2006nematic} for passive nematics. 
The directions of the $+1/2$ defects are determined by the azimuthal $m$-fold symmetries. 
Around these $+1/2$ defects, active stresses are large and create a directional flow, so that regions with large velocity magnitudes coincide with the vicinity of the $+1/2$ defects. 
Finally, as the degree $l$ increases, finer-scale structures appear along the latitudinal direction, e.g., octopolar for $l=3$ as shown in figure \ref{fig:eigenmodes}($b$).

\section{Numerical method} \label{sec:num method}

We apply a pseudo-spectral method to solve the full nonlinear system based on the SWSH expansion of the distribution function \eqref{eq:Psinequation} and its relation to the velocity field \eqref{eq:uPDFcoeff}, where the nonlinear products are performed in real space and the linear coefficients are calculated in spectral space. 
We apply a second-order semi-implicit time marching scheme (with explicit Euler for the very first time step) \citep{tornberg2004simulating}. Specifically, we discretize the components of the Fokker-Planck equation \eqref{eq:Psinequation} as
\begin{align}
\begin{split}
    \frac{1}{2\Delta t}(3 a_{lm}^{(-n), t+1} - 4a_{lm}^{(-n),t} + a_{lm}^{(-n),t-1}) = D_{ln} a_{lm}^{(-n),t+1} & + U_{ln}^- a_{lm}^{(-n+1),t+1} + U_{ln}^+ a_{lm}^{(-n-1),t+1} \\
    & + 2 N(a_{lm}^{(-n),t}) - N(a_{lm}^{(-n), t-1}),
\end{split}
\end{align}
where the implicit linear coefficients are 
\begin{align}
    & D_{ln} = -D[l(l+1)-n^2] - n^2, \\
    &  U^-_{ln} = - \frac{U}{2}\sqrt{(l-n+1)(l+n)} ,\\
    & U^+_{ln} = \frac{U}{2}\sqrt{(l+n+1)(l-n)},
\end{align}
and the explicit nonlinear parts contain the SWSH coefficients of 
\begin{align}
    & \frac{1}{\sqrt{2}} (u^+\bar{\eth}\Psi_n + u^- \eth \Psi_n) 
    + \frac{n}{2\sqrt{2}}
   [\gamma(\eth u^+ \Psi_{n+2}
    - \bar{\eth} u^- \Psi_{n-2}) 
    - 2\eth u^- \Psi_n], 
\end{align}
which are calculated using the forward SWSH transforms. 
For the forward and backward SWSH transforms,  we use a Python package with the algorithm introduced in \cite{price2024differentiable}. 
The number of SWSH modes is $L=96,\; N=95$ on a Gauss-Legendre mesh for all simulations. 
To prevent aliasing errors, we also apply a dealiasing method with up-sampling band limit $L_{\text{up}} = 3L/2$ \citep{beyer2014numerical}. 
Note that dealiasing in $N$ is not necessary since there is no product in real space for the $n$-components.

\section{Results and discussion} \label{sec:num results}
\subsection{Nonlinear dynamics and relation between polarity and nematic fields}

\begin{figure}
\centerline{\includegraphics[width=\textwidth]{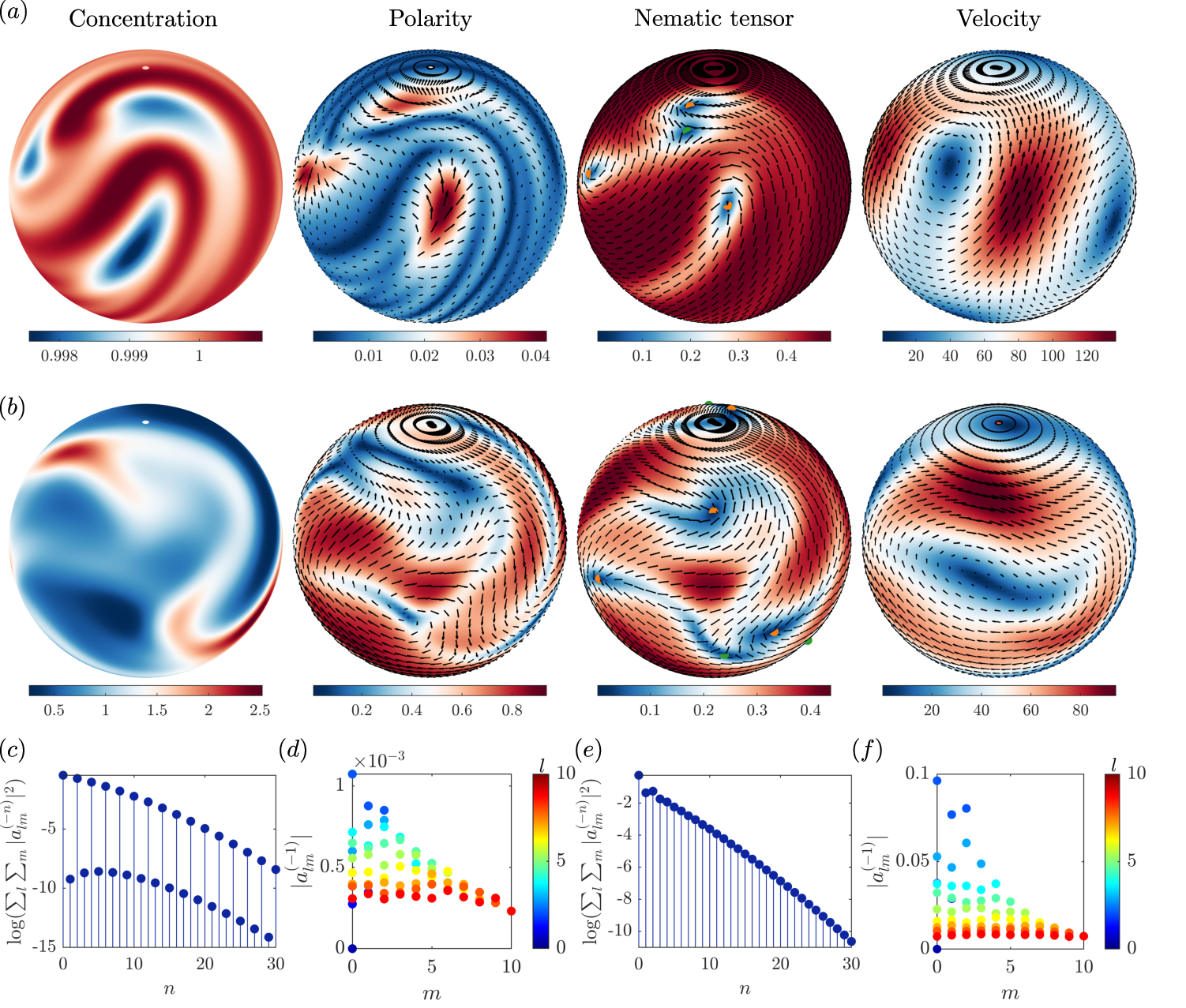}}
  \caption{($a,b$) Snapshots of the concentration, polarity, nematic tensor, and velocity fields in the nonlinear regime in two typical simulations. Green and orange dots marked on the nematic field correspond to $+1/2$ and $-1/2$  defects. Also see videos 1 and 2 of the Supplementary Material. Bottom panels show the corresponding plots of the time-averaged $N$-spectrum ($c,e$), and time-average of the first $10$ spin-weighted spherical harmonic coefficients of the polarity field ($d,f$) in the fully developed regime. Parameter values are
 $\alpha = -500$, $U=1$, $D=1$, 
 $\lambda=0.01$ in ($a,c,d$), and  
   $\alpha = -500$, $U=100$, $D=1$, 
   $\lambda=0.01$ in ($b,d,f$). 
   Note the small magnitude of concentration fluctuations and of the polarity when $U$ is small in case ($a$). }
\label{fig:contours}
\end{figure}

We perform nonlinear numerical simulations starting from small random perturbations around the uniform isotropic state. 
The fluctuations grow due to the linear instability, followed by nonlinear, chaotic regimes where the energy injection due to the nematic active stress is balanced by the damping effect of fluid viscosity, rotational and translational diffusion. 
As a validation of the numerical method, the growth rates from numerical simulations in the linear regime were verified to agree well with the predictions from the linear stability analysis. 

Typical snapshots of the concentration, polarity, nematic tensor, and velocity fields recovered from the probability distribution function are shown in figure \ref{fig:contours}($a$,$b$) for two cases corresponding to a small ($U=1$, panel $a$) and large ($U=100$, panel $b$) self-propulsion speed (also see videos 1 and 2 of the Supplementary Material). 
The concentration, polarity, nematic, and velocity fields exhibit spatial fluctuations, characterized by the formation and breakup of high-concentration and high-polarity regions, together with the emergence and annihilation of $+1/2$ and $-1/2$ defects. 
Note that due to the topology of the sphere, the sum of all defect charges in the nematic field is always $2$. 
In the small self-propulsion speed regime (figure \ref{fig:contours}$(a,c)$), the concentration fluctuations are weaker than in the large self-propulsion case, and the polarity magnitudes are also smaller. 
The reason for this can be seen from equation \eqref{eq:Psinequation}: the self-propulsion term couples the closest neighboring orientational moments. 
This is also verified in the spectrum of the spin weight $N$, summed over $l$ and $m$ ($N$-spectrum), which exhibits alternating magnitudes between successive moments, while in the large self-propulsion speed regime (figure \ref{fig:contours}$(b,e)$), the magnitudes of neighboring moments differ less.  
Also, we notice that the polarity field contains more banded spatial structures in the small self-propulsion speed regime, whereas it is more spatially isotropic in the large self-propulsion regime. 
This can also be seen in their spectra in figure \ref{fig:contours}$(d,f)$: the magnitudes of the coefficients are more equally distributed for all modes in the small self-propulsion speed case, while they become more concentrated at low values of $m$ and $l$ as the self-propulsion speed increases.

\begin{figure}
\centerline{\includegraphics[width=\textwidth]{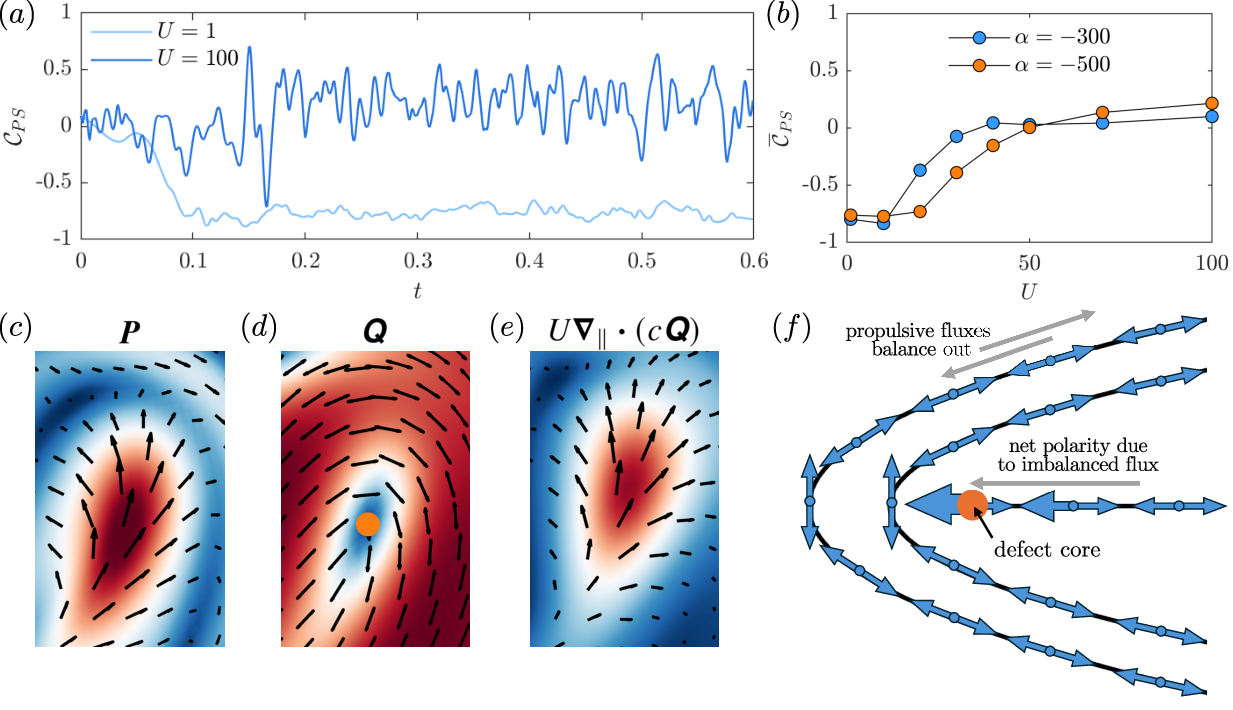}}
  \caption{($a$) Temporal evolution of the correlation $\mathcal{C}_{PS}(t)$ between the magnitude $P$ of the polarity and the nematic scalar order parameter $S$, as defined in equation (\ref{eq:correlationPS}), in the low and high propulsion regimes: $U=1$ and $100$. Other parameters are: $\alpha = -500$, $D=1$, and 
  $\lambda=0.01$. 
  ($b$) Time-averaged correlation $\smash{\overline{\mathcal{C}}_{PS}}$ in the fully developed regime as a function of the self-propulsion speed $U$, for two different activity levels $\alpha$. 
  ($c,d, e$)~Zoom-ins of the polarity field $\boldsymbol{P}$, nematic tensor field $\mathsfbi{Q}$, and coupling field $U\bnabla_{\parallel}\bcdot(c\mathsfbi{Q})= U(\eth \Psi_2 \mm + \bar{\eth}\Psi_{-2} \bmm)/2$ in the vicinity of a $+1/2$ nematic defect (orange dot) for $\alpha = -500$, $U=1$, $D=1$, and 
  $\lambda=0.01$. See the colorbars in figure \ref{fig:contours}($a$). 
  ($f$) Schematic illustrating the mechanism of polarity generation near $+1/2$ nematic defects as a result of self-propulsion. 
   }
\label{fig:correlation}
\end{figure}

In both cases, $+1/2$ defects in the nematic tensor field correlate with regions of high velocity in the flow field: indeed the sharp gradients that occur in the nematic field near these defects result in strong active forces in the surface Stokes equation. 
In the case of weak self-propulsion, we observe that the positions of the defects in the nematic tensor field also coincide with regions of high polarity (figure \ref{fig:contours}$(a,b)$). 
No such correlation is observed in the large self-propulsion regime. 
To more quantitatively characterize this effect, we plot in figure \ref{fig:correlation}($a$) the temporal evolution of the correlation between the magnitude $P$ of the polarity field and the nematic scalar order parameter $S$ (defined as the maximum eigenvalue of the nematic tensor):
\begin{equation}
    \mathcal{C}_{PS}(t)=\frac{\int_{S^2} (P-\langle P\rangle)(S-\langle S\rangle)\dd \Omega}{\sqrt{\int_{S^2} |P-\langle P\rangle|^2 \dd \Omega}\sqrt{\int_{S^2} |S-\langle S\rangle|^2\dd \Omega}}. \label{eq:correlationPS}
\end{equation} 
In the regime of weak self-propulsion ($U=1$), there is a clear anti-correlation between these two fields, reflecting the relationship between defect positions of the nematic field and the high polarity regions, while no such correlation is observed when self-propulsion is strong ($U=100$). 
The time-averaged autocorrelation function in the fully developed regime is plotted versus $U$ in figure \ref{fig:correlation}($b$) for activity $\alpha=-300$ and $\alpha = -500$, where a transition from the anti-correlated state to the uncorrelated state is observed as $U$ increases. 
The critical self-propulsion speed $U$ for the transition to occur decreases as the activity $\alpha$ decreases. 

To elucidate the origins of the correlation observed at low $U$, we zoom in on the details of the polarity and nematic fields close to a $+1/2$ defect in figure \ref{fig:correlation}($c,d$). While polarity is negligible along the nematic field lines that wrap around the defect front, it is significant along the axis of the defect where it points into the direction of the bend. A qualitative mechanism for this coupling is illustrated in figure \ref{fig:correlation}($f$). Along nematic field lines, particles with orientations that are diametrically opposed along the director field swim in opposite directions. Away from defects, this results in two opposing particle fluxes that nearly cancel each other out. Along the axis of a $+1/2$ defect, however, the propulsive fluxes into the defect and away from it become imbalanced, generating a net polarity in that direction. This effect is generic in systems of nematically aligning self-propelled particles \citep{baskaran2012}.  Mathematically, this coupling arises from the terms $U(\eth \Psi_2)/2$ and $U(\bar{\eth}\Psi_{-2})/2$ in the moment hierarchy equations (\ref{eq:Psinequation}), which corresponds to a source term $U\bnabla_\parallel\bcdot (c\mathsfbi{Q})$ in the evolution equation of the polarity stemming from the interplay of self-propulsion and nematic field gradients. That source term, plotted in figure \ref{fig:correlation}($e$), indeed mirrors the polarity field near the defect core. This net polarity at $+1/2$ defects adds to their forward motion, which is primarily driven by the strong active flow they generate (figure \ref{fig:correlation}($a$)).
No such structure is observed in the high self-propulsion regime ($U=100$, figure \ref{fig:contours}($b$)), which we attribute to two effects: first, we find that sharply defined $+1/2$ defects are absent and instead replaced by elongated defect bands; second, strong self-propulsion has a smoothing effect on the spatial features of the orientational moments as it tends to mix orientations \citep{albritton2023}.  

\subsection{Power spectra and energy fluxes}
\begin{figure}
\centerline{\includegraphics[width=\textwidth]{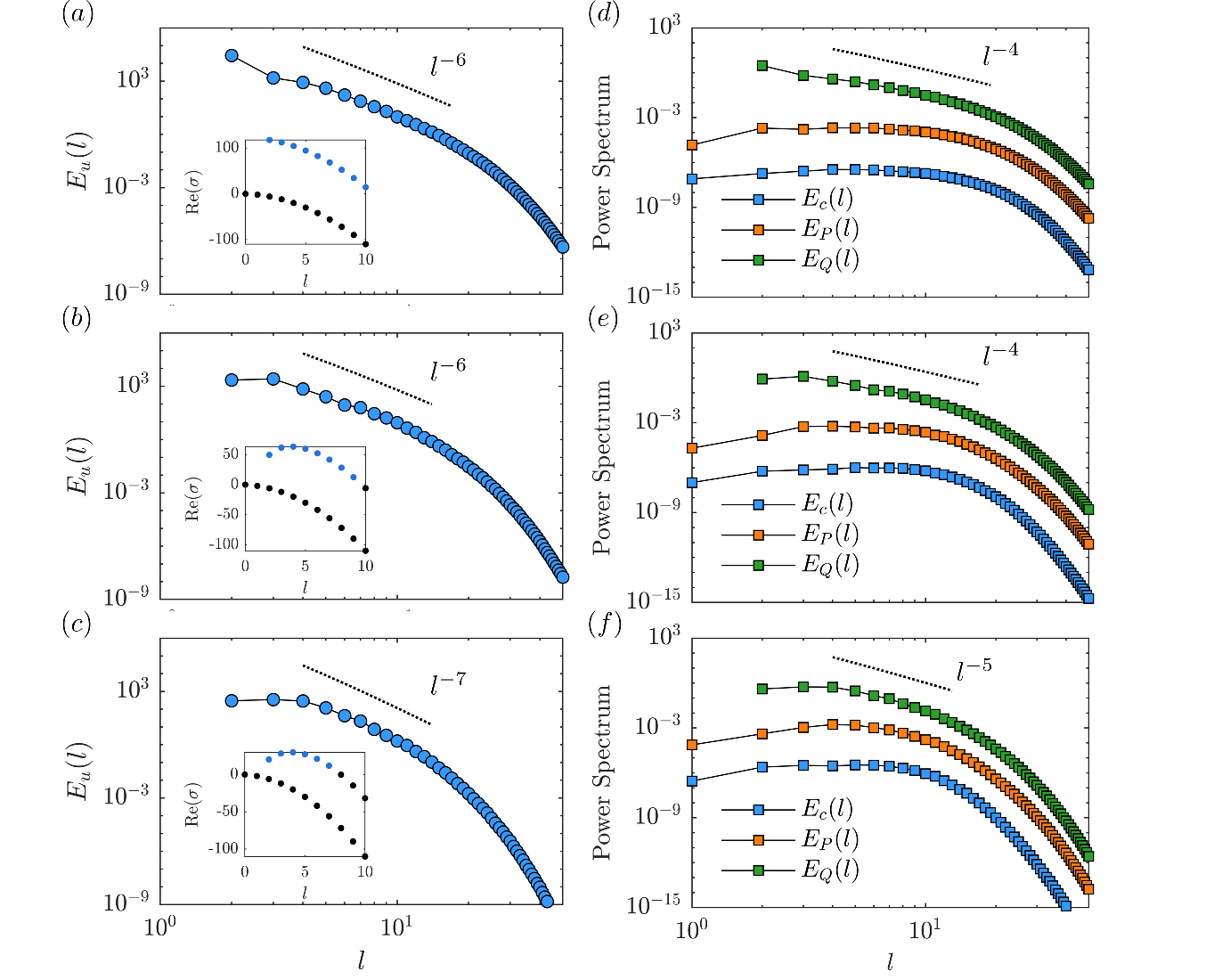}}
  \caption{($a$-$c$) Time-averaged kinetic energy spectra, defined in equation (\ref{eq:energyspectrum}), for increasing values of $\lambda$. Insets show the corresponding real parts of the growth rates from linear stability analysis. ($d$-$f$)~Time-averaged energy spectra of the concentration, polarity and nematic fields, defined in equations (\ref{eq:cspectrum})--(\ref{eq:Qspectrum}). Parameters values: ($a,d$) $\alpha = -500$, $U=1$, $D=1$, 
  $\lambda=0.01$;
  ($b,e$) $\alpha = -500$, $U=1$, $D=1$, 
  $\lambda=1$; ($d,f$) $\alpha = -500$, $U=1$, $D=1$, 
  $\lambda=3$. 
  The scaling factors are fitted from the data and rounded to the nearest integer. }
\label{fig:l-spectrum}
\end{figure}

Upon increase of the relative viscosity, viscous damping in the bulk becomes more significant and results in mode selection. 
To characterize the impact of this effect on turbulent structures in the nonlinear regime, we introduce the kinetic energy and the $L^2$ norm, or effective energy, of the concentration, polarity, and nematic fields as
\begin{align}
    & E_u = \frac{1}{2}\int_{S^2} |\mbf{u}|^2 \dd \Omega = \sum_{l=1}^\infty \sum_{m=-l}^l |u_{lm}^+|^2 \coloneqq \sum_{l=1}^\infty E_u(l), \label{eq:energyspectrum}\\
    & E_c = \int_{S^2} |c|^2 \dd \Omega = 4\pi^2 \sum_{l=0}^\infty \sum_{m=-l}^l |a_{lm}^{(0)}|^2 \coloneqq \sum_{l=0}^\infty E_c(l), \label{eq:cspectrum}  \\
    & E_P = \int_{S^2} |c\mbf{P}|^2 \dd \Omega = 4\pi^2 \sum_{l=1}^\infty \sum_{m=-l}^l |a_{lm}^{(-1)}|^2 \coloneqq \sum_{l=1}^\infty E_P(l),\\
    & E_Q = \int_{S^2} |c\mbf{Q}|^2 \dd \Omega = 2\pi^2 \sum_{l=2}^\infty \sum_{m=-l}^l |a_{lm}^{(-2)}|^2 \coloneqq \sum_{l=2}^\infty E_Q(l). \label{eq:Qspectrum}
\end{align}
Time-averaged kinetic energy spectra and moment energy spectra in the small self-propulsion speed regime $(U=1)$ are shown in figure \ref{fig:l-spectrum}. 
As predicted by the real parts of the growth rates from linear stability analysis (see insets in figure \ref{fig:l-spectrum}($a$-$c$)), the most unstable modes shift to higher wavenumbers as the relative viscosity increases. 
This mode selection is also observed in the nonlinear regime, where the time-averaged kinetic energy spectra as well as the spectra of the moments are peaked at higher wavenumbers for larger relative viscosities. This indicates that the energy injection in these regimes occurs at this length scale, and is consistent with the emergence of  finer-scale structures in the flow field. 
In addition, we find that the spectra at high wavenumbers decay faster for larger relative viscosities. 

The energy spectra also indicate a potential scaling law at low wavenumbers. 
For small and intermediate relative viscosity (figure \ref{fig:l-spectrum}($a$,$b$,$d$,$e$)), the kinetic energy spectra show a $l^{-4}$ power law and the nematic spectra show a $l^{-6}$ power law. 
The difference between the scaling exponents of these two spectra can be explained by the relation between the coefficients of the velocity field and the coefficients of the second moments in the SWSH expansions. 
From the force balance at the interface, we have 
\begin{align}
    \langle |u_{lm}^+|^2 \rangle = \Big(\frac{\pi\alpha}{2\sqrt{2}s_l}\Big)^2(l-1)(l+2) \langle |a_{lm}^{(2)} - a_{lm}^{(-2)}|^2 \rangle. 
\end{align}
In the limit where $\lambda$ is small, the coefficient $s_l$ scales as $l^2$. 
Therefore, there exists an $l^{-2}$ difference between the velocity and the nematic energy spectra. 
A difference also persists as the relative viscosity increases (figure \ref{fig:l-spectrum}($e$,$f$)), although the energy spectra are found to decay faster in this case. 
Note that the scalings observed here differ from the universal scaling law reported previously in flat active nematic interfaces \citep{martinez2021scaling} due to the absence of nematic elasticity in the present model. 

To further understand the energy transfer occurring between scales in the fully developed turbulent regime, we analyze the time-averaged energy spectral equation of the $\Psi_2$ mode, namely, the nematic order parameter field, which reads 
\begin{align}
    \pdif{E_2(l)}{t} =  I_{2,\text{sp}}(l) + D_2(l) + T_{2,\text{adv}}(l) + I_{2,\text{fa}}(l) + T_{2,\text{coro}}(l).  \label{eq:E2eq}
\end{align}
Here we have introduced the energy of the $\Psi_2$ mode as
\begin{align}
 E_2(l) = \frac{1}{4\pi^2} \langle E_Q(l) \rangle = \frac{1}{2} \sum_{m=-l}^l \langle|a_{lm}^{(-2)}|^2 \rangle.
\end{align}
There are five contributions to the energy spectral equation: the energy source/sink $I_{2,\mathrm{sp}}$ due to self-propulsion; the damping $D_2$ due to translational and rotational diffusion; the energy transfer $T_{2,\text{adv}}$ due to fluid advection; the power injection $I_{2,\text{fa}}$ due to flow-alignment; and the energy transfer $T_{2,\text{coro}}$ due to the corotational term in Jeffery's equation. 
Among them, the diffusion term is calculated directly as
\begin{align}
    D_2(l) = - \sum_{m=-l}^l [D(l^2+l-4)+4] \langle |a_{lm}^{(-2)}|^2 \rangle,
\end{align}
while the coefficients of the other nonlinear terms are calculated numerically in real space, then transformed to the SWSH spectral space as
\begin{align}
    \mathcal{N}(l) = \sum_{m=-l}^l \langle \text{Re}\{\bar{a}_{lm}^{(-2)} \mathcal{N}_{lm}\} \rangle. 
\end{align}
We also introduce the total energy flux as the cumulative sum of each contribution term for modes $l\le k$, for example:
\begin{align}
    \Pi_{T}(k) = \sum_{l=0}^{k}T(l),\quad \Pi_{I}(k) = \sum_{l=0}^{k}I(l), \quad \Pi_{D}(k) = \sum_{l=0}^{k}D(l). \label{eq:Piterms}
\end{align}
The total energy flux is positive for an energy source, and negative for an energy sink. A zero total energy flux indicates there is no net energy injection or dissipation and only energy transfer exists. 

\begin{figure}
\centerline{\includegraphics[width=\textwidth]{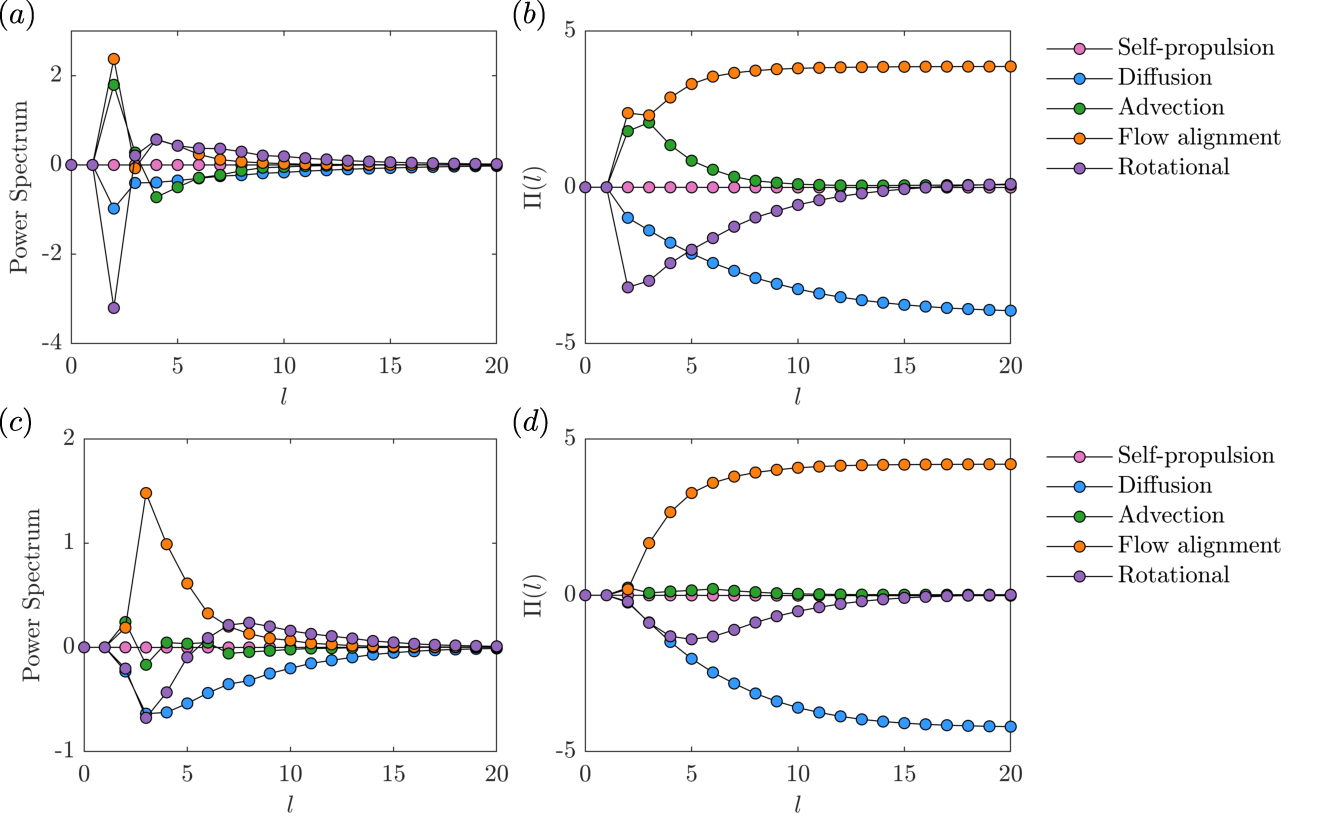}}
  \caption{($a$,$c$) Time-averaged contributions, at small $U$, to the energy spectral equation (\ref{eq:E2eq}) of the nematic field, and ($b$,$d$) corresponding cumulative energy fluxes defined in equation (\ref{eq:Piterms}). Parameter values:  
  ($a,b$) $\alpha = -500$, $U=1$, $D=1$, 
  $\lambda=0.01$;
   ($c,d$)~$\alpha = -500$, $U=1$, $D=1$, 
   $\lambda=1$.   }
\label{fig:energy flux nematic small U}
\end{figure}

Figure \ref{fig:energy flux nematic small U} shows the contributions of each term to the energy spectral equation and the total energy fluxes in the low self-propulsion speed regime ($U=1$) for $\lambda=0.01$ and $\lambda=1$. 
In both cases, the flow-alignment terms contribute to the power injection, which is balanced by the dissipation due to translational and rotational diffusion. 
The self-propulsion terms are approximately zero at all length scales, indicating that there is almost no energy injected or dissipated due to this effect. 
The advection term acts like an energy source at large length scales and like an energy sink at small length scales in the low relative viscosity case, indicating energy transfer from small length scales to large length scales due to advection. 
This active inverse energy cascade is reminiscient of the inverse energy cascade in two-dimensional inertial turbulence, and is less pronounced as the bulk viscosity increases (increasing relative viscosity). 
In contrast, the rotational terms in Jeffery's equation transfer energy from large length scales to small length scales in both cases. 
This differs from the active nematic case where there is no energy cascade \citep{alert2020universal}. 
Additionally, as the relative viscosity increases, the energy transfer due to the advection term becomes weaker, and the peak of the energy injection in the flow-alignment term shifts to higher wavenumbers (figure \ref{fig:energy flux nematic small U}($c$,$d$)), consistent with the mode selection predicted in the linear stability analysis and exhibited in the energy spectra. 

\begin{figure}
\centerline{\includegraphics[width=\textwidth]{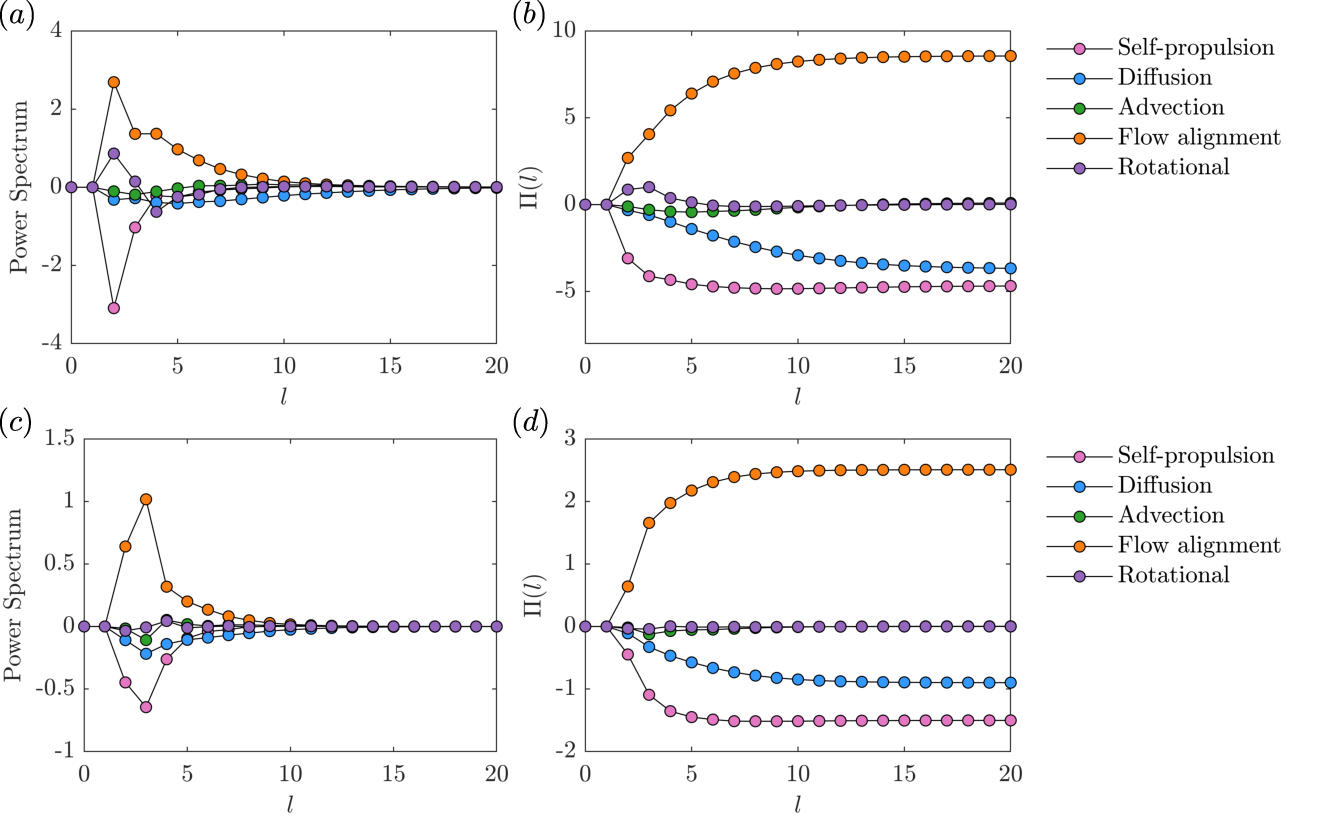}}
  \caption{($a$,$c$) Time-averaged contributions, at large $U$, to the energy spectral equation (\ref{eq:E2eq}) of the nematic field, and ($b$,$d$) corresponding cumulative energy fluxes defined in equation (\ref{eq:Piterms}). Parameter values:  
  ($a,b$) $\alpha = -500$, $U=100$, $D=1$, 
  $\lambda=0.01$;
   ($c,d$) $\alpha = -500$, $U=100$, $D=1$, 
   $\lambda=1$.   }
\label{fig:energy flux nematic large U}
\end{figure}

\begin{figure}
\centerline{\includegraphics[width=\textwidth]{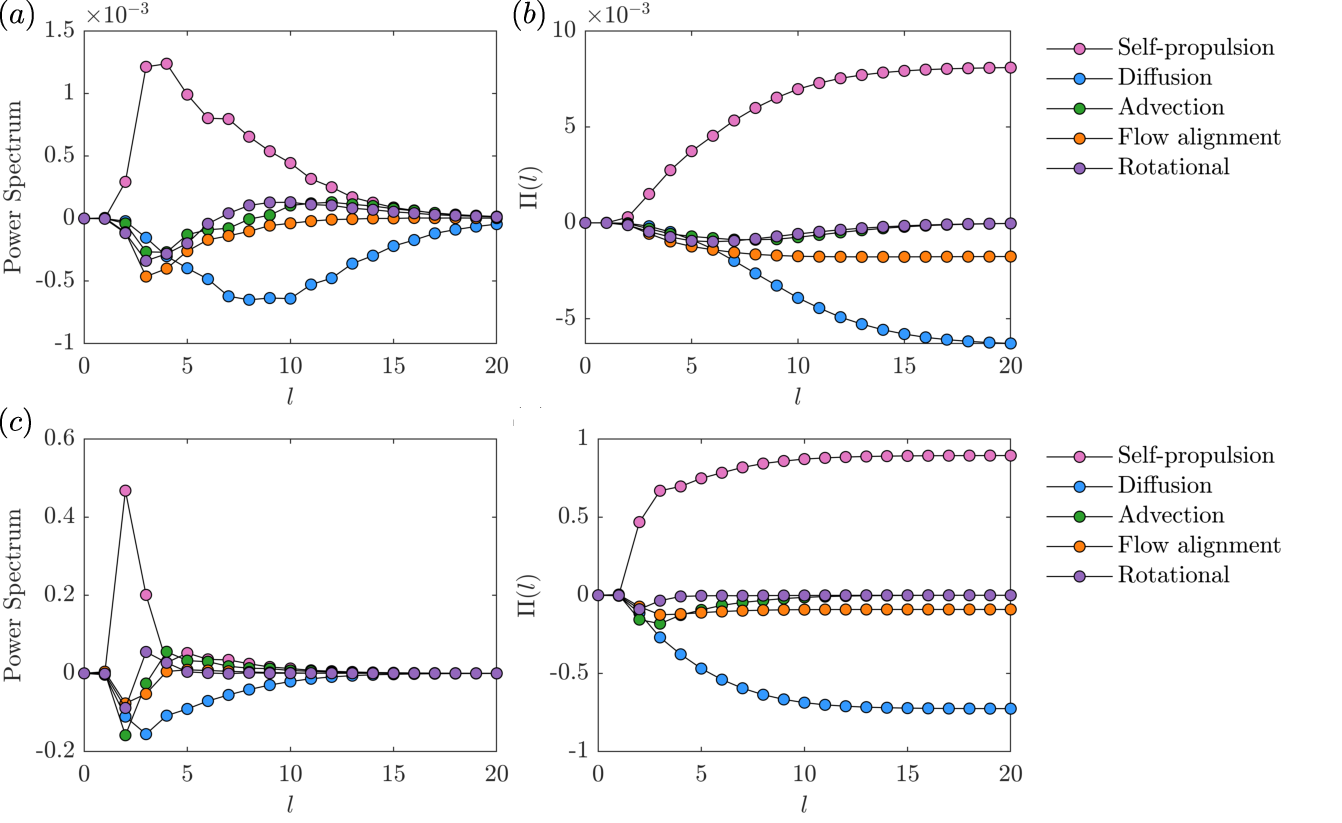}}
  \caption{($a$,$c$) Time-averaged contributions to the energy spectral equation of the polarity field, and ($b$,$d$) corresponding cumulative energy fluxes. Parameter values:  
  (a,b) $\alpha = -500$, $U=1$, $D=1$, 
  $\lambda=1$;
   ($c,d$) $\alpha = -500$, $U=100$, $D=1$, 
   $\lambda=1$.   }
\label{fig:energy flux polarity}
\end{figure}

The self-propulsion term in the energy spectral equation plays a more important role as the self-propulsion speed increases (figure \ref{fig:energy flux nematic large U}, $U=100$), while the energy transfer due to advection and the rotational term in Jeffery's equation becomes less significant. 
The energy transfers due to the advection and the corotational term, though very small, change their direction compared to the small self-propulsion speed case. 
In addition, the self-propulsion term behaves as an effective energy sink in the spectral equation for the nematic field. 
This suggests that the energy removed by this term is transferred to other moments. 
To verify this assumption, we further plot in figure \ref{fig:energy flux polarity} the time-averaged contributions and total energy fluxes in the energy spectral equation for the polarity field (defined analogously to equation (\ref{eq:E2eq}) but using the $\Psi_1$ mode) for both $U=1$ and $U=100$. 
For the polarity field, instead of the flow-alignment, it is the self-propulsion term that injects energy, while flow alignment acts as an energy sink. 
The energy injection from the self-propulsion term is nearly negligible when the self-propulsion speed is small (figure \ref{fig:energy flux polarity}($a$,$b$)), and its magnitude increases as the self-propulsion speed increases (figure \ref{fig:energy flux polarity}($c$,$d$)). 
In summary, the contributions to the energy spectral equation and energy fluxes demonstrate that energy is injected into the system via the flow-alignment term in the nematic field, and transferred into other moments as a result of self-propulsion.

\subsection{Entropy production rate}
Finally, to characterize the spatial inhomogeneity of the system, we introduce the total configurational entropy as
\begin{align}
    S(t) = \int_{\mathcal{M}} \dd \mbf{x} \int_{S^1} \dd \psi \frac{\Psi}{\Psi_0} \ln\Big(\frac{\Psi}{\Psi_0}\Big), \label{eq:entropy}
\end{align}
which quantifies departures from the uniform isotropic state $\Psi_0$ \citep{saintillan2008instabilities}. 
After taking the time derivative and using the conservation equation, we obtain the entropy production rate, which is formally identical to the flat case, and reads
\begin{align}
    \dot{S}(t) = - \int_{\mathcal{M}} \dd\mbf{x} \int_{S^1} \dd \psi (D|\nabla_x \ln\Psi|^2 + |\partial_\psi \ln\Psi|^2)
    + \frac{2\beta}{\alpha}\int_{\mathcal{M}}  \mbf{\Sigma}^{a} \boldsymbol{:} \mathsfbi{E} \dd \mbf{x}.
\end{align}
Similar to the flat case, the first term on the right-hand side arising from translational and rotational diffusion  can only be negative, and thus drives the system towards isotropy. The last term, however, captures the effects of active stresses and can be positive, driving the system away from equilibrium. This term can be interpreted as the power input injected into the fluid by active stresses, and can be related via a mechanical energy balance to the total rate of viscous dissipation in the system.   
Indeed, multiplying the surface momentum balance equation by $\mbf{u}$, integrating over the surface and performing integration by parts over the entire fluid domain yields
\begin{align}
\begin{split}
    \int_{\mathcal{M}}  \mbf{\Sigma}^{a} \boldsymbol{:} \mathsfbi{E} \dd \mbf{x} 
    & = - 2 \int_{\mathcal{M}}  \mathsfbi{E} \boldsymbol{:} \mathsfbi{E} \dd \mbf{x} 
    - \int_{V^{\text{out}}} {\bnabla}\bcdot(\mbf{\sigma}^{\text{out}}\bcdot\mbf{u}) \dd \mbf{x} 
    - \int_{V^{\text{in}}} {\bnabla}\bcdot(\mbf{\sigma}^{\text{in}}\bcdot\mbf{u}) \dd \mbf{x} \\
    & = - 2 \int_{\mathcal{M}}  \mathsfbi{E} \boldsymbol{:} \mathsfbi{E} \dd \mbf{x} 
    - 2\int_{V^{\text{out}}}  \lambda^{\text{out}}\tilde{\mathsfbi{E}}\boldsymbol{:}\tilde{\mathsfbi{E}} \dd \mbf{x} 
    - 2 \int_{V^{\text{in}}} \lambda^{\text{in}}\tilde{\mathsfbi{E}}\boldsymbol{:}\tilde{\mathsfbi{E}} \dd \mbf{x}, 
    \end{split}\label{eq:energybudget}
\end{align}
where $\tilde{\mathsfbi{E}}$ denotes the bulk strain-rate tensor. 
The three terms on the right-hand side describe the rates of viscous dissipation in the membrane and in the bulk fluids and can only be negative. 
Hence, the entropy production rate can be positive only in pusher suspensions ($\alpha < 0$). 

\begin{figure}
\centerline{\includegraphics[width=0.95\textwidth]{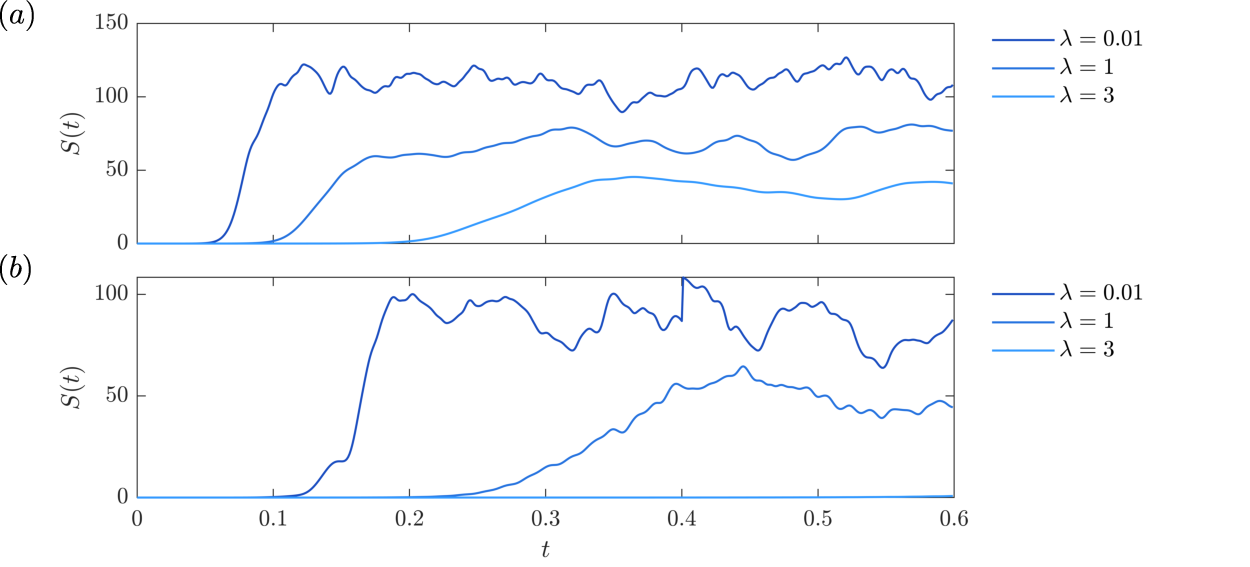}}
  \caption{Temporal evolution of the total configurational entropy defined in equation (\ref{eq:entropy}) for different relative viscosities. Parameter values: 
  $(a)$ $\alpha = -500$, $U=1$, $D=1$; 
  $(b)$ $\alpha = -500$, $U=100$, $D=1$.}
\label{fig:entropy}
\end{figure}

\begin{figure}
\centerline{\includegraphics[width=\textwidth]{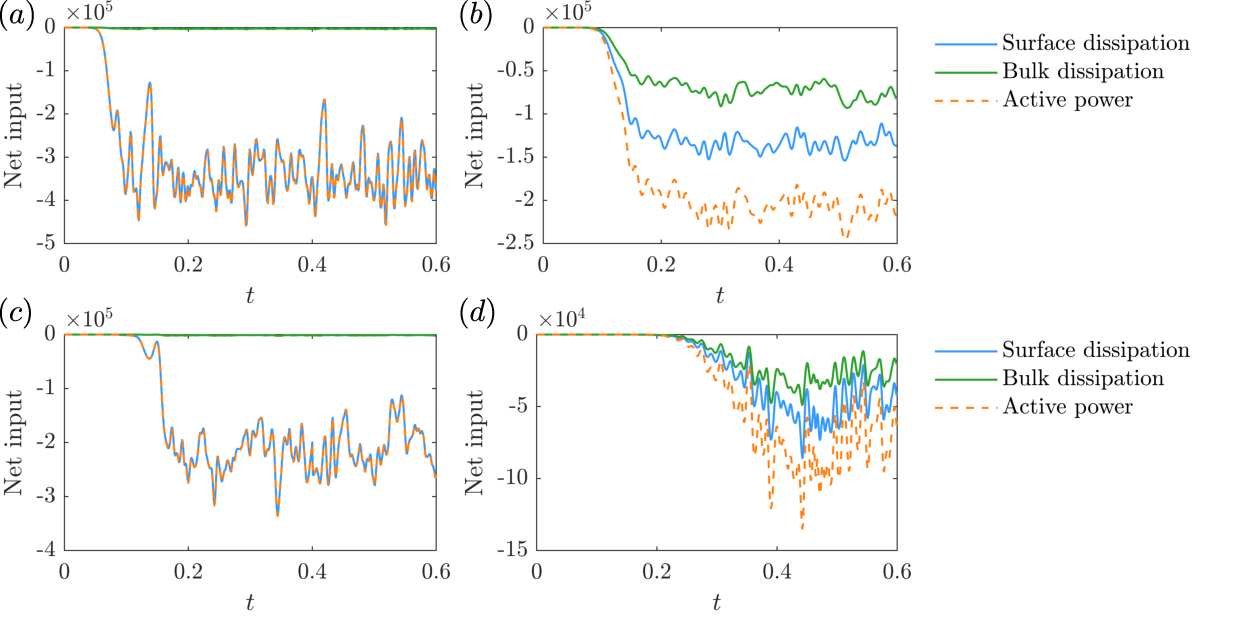}}
  \caption{Temporal evolution of the various terms in the mechanical energy balance of equation (\ref{eq:energybudget}). Parameter values: 
  $(a)$ $\alpha = -500$, $U=1$, $D=1$, 
  $\lambda=0.01$;
  $(b)$~$\alpha = -500$, $U=1$, $D=1$, 
  $\lambda=1$;
  $(c)$ $\alpha = -500$, $U=100,\;D=1$, 
  $\lambda=0.01$;
  $(d)$ $\alpha = -500 $, $U=100$, $D=1$, 
  $\lambda=1$.}
\label{fig:entropy terms}
\end{figure}

The temporal evolution of the total entropy for suspensions of pushers in the regimes of small and large self-propulsion speeds are plotted in figure \ref{fig:entropy} for various relative viscosities. 
In both regimes, the saturation time of the total entropy increases as the relative viscosity increases, indicating slower dynamics due the additional damping occurring in the bulk. 
The magnitude of total entropy is also found to be smaller for larger relative viscosities, despite the emergence of fluctuations on smaller length scales. 
The reason for this is that there is more dissipation due to viscosity, resulting in slower dynamics.
This can be seen in figure \ref{fig:entropy terms}, which shows the time evolution of the various contributions in the mechanical energy budget of equation (\ref{eq:energybudget}). 
While the contribution from the bulk dissipation rate is found to increase when the relative viscosity increases, the surface dissipation rate significantly decreases, causing the total dissipation rate, and therefore the magnitude of the active power, to decrease overall. 
As the self-propulsion speed increases (figure \ref{fig:entropy}$(b)$ and figure \ref{fig:entropy terms}$(c,d)$), the total entropy, as well as the various contributions of the energy budget, decrease in magnitude as advection by self-propulsion tends to damp inhomogeneities driven by the system's instability.

\section{Concluding remarks} \label{sec:conclusion}
We have presented a kinetic model and simulations for the collective dynamics of a dilute suspension of rod-like active particles confined to a curved viscous interface. 
Our analysis focused on the case of a spherical geometry. 
Using spin-weighted spherical harmonics as a natural basis to analyze this system, we performed a linear stability analysis around the uniform, isotropic state. 
In contrast to the case of bulk suspensions which are characterized  by a long-wavelength instability, the relative viscosity, or equivalently the ratio between the radius of the sphere and Saffman-Delbr\"uck length, was shown to introduce a mode selection effect. 
Our theoretical analysis was complemented by numerical simulations using a pseudo-spectral method based on spin-weighted spherical harmonics, which additionally demonstrated a correlation between the polarity and the nematic order parameter fields at low self-propulsion speeds. 
The numerical energy spectra of the kinetic energy and nematic order parameter both showed qualitative agreement with the mode selection effect predicted by the linear analysis. 
A closer analysis of the energy spectral equation revealed that the flow-alignment term in the nematic field acts as source of fluctuations, while self-propulsion results in energy transfer between orientational moments. 
In the regime of low self-propulsion speeds, an inverse energy cascade driven by the advection term was observed. 
Finally, we found that the total entropy decreases as the relative viscosity and the self-propulsion speed increase, and that the contribution of the bulk viscous dissipation to the entropy production rate increases as the relative viscosity increases. 

Extensions of the present study may consider the dynamics of active suspensions on arbitrarily curved, static or evolving interfaces. 
In these cases, the presence of multiple geometric length scales, as well as the interplay between the particle distribution and spatially varying curvature, could potentially give rise to further rich phenomena. 
One potential way to address the associated numerical challenges is to combine the boundary integral method \citep{pozrikidis1992boundary} with the discrete $\eth$ operators on a triangular mesh. 
A related approach was recently proposed in \cite{zhu2025active} for active nematodynamics on Riemannian 2-manifolds within the framework of the complex line bundle. 
Yet, the Fokker-Planck equation and its coupling to bulk fluid flows have not been considered. 
Another potential approach is to use conformal mappings to pull back the dynamics on a general surface with spherical topology to $S^2$, following the method in \cite{beyer2014numerical}. 
The resulting dynamics could then be analyzed within the spectral framework on $S^2$. 


\backsection[Acknowledgements]{We thank Albert Chern for useful discussions on the spin-weighted spherical harmonics. }

\backsection[Funding]{D.S. acknowledges funding from National Science Foundation grant No.~DMS-2153520.}

\backsection[Declaration of interests]{The authors report no conflict of interest.}

\backsection[Data availability statement]{The simulation software is available from the authors upon reasonable request. }

\backsection[Author ORCIDs]{Y. Chen, https://orcid.org/0009-0003-4564-6606, V.P. Patil, https://orcid.org/0000-0002-0815-6646, D. Saintillan, https://orcid.org/0000-0001-9948-708X}

\backsection[Author contributions]{All authors designed the research. Y.C. developed the theoretical model, performed the linear stability analysis, wrote the simulation software, performed simulations, analyzed data, prepared figures and wrote the manuscript. All authors interpreted results and data. V.P. and D.S. edited the manuscript.}

\appendix

\section{Basic notations in differential geometry and derivation of the flux terms in coordinate form} \label{appA}
Given a 2D Riemannian surface embedded in $\mathbb{R}^3$ denoted by $(y^1,y^2)\mapsto \mbf{x}\in \mathcal{M} \subset \mathbb{R}^3$, the covariant basis and the normal vector form a local basis:
\begin{align}
    \mbf{g}_1 = \pdif{\mbf{x}}{y^1},\quad \mbf{g}_2 =  \pdif{\mbf{x}}{y^2}, \quad \mbf{N} = \frac{\mbf{g}_1\times\mbf{g}_2}{|\mbf{g}_1\times\mbf{g}_2|}. 
\end{align}
A tensor in the tangent plane can then be expressed as $\mathsfbi{T} = T_{\alpha}^{\;\beta} \mbf{g}^\alpha \otimes \mbf{g}_\beta$, where we have used the Einstein summation convention, with Greek indices varying from $1,2$. 
With the covariant components of the metric tensor $g_{\alpha\beta} = \mbf{g}_\alpha \bcdot \mbf{g}_\beta$ and the contravariant components $(g^{\alpha\beta}) = (g_{\alpha\beta})^{-1}$, we can lower or raise the index of a tensor component as $T_{\alpha\beta} = g_{\alpha\gamma} T^{\gamma}_{\;\beta}$ and $T^{\alpha\beta} = g^{\alpha\gamma} T_{\gamma}^{\;\beta}$. 

The covariant derivative of a vector field in the tangent plane $\mbf{v} = v^\alpha \mbf{g}_\alpha$ is denoted by
\begin{align}
    \nb_\parallel \mbf{v} = \nabla_\alpha v^\beta \mbf{g}^\alpha \otimes \mbf{g}_\beta, 
\end{align}
where
\begin{align}
    \nabla_\alpha v^\beta = \pdif{v^\beta}{y^\alpha} + \Gamma_{\alpha \gamma}^\beta v^\gamma,
\end{align}
with the Christoffel symbols, $\Gamma$, defined by
\begin{align}
    \Gamma_{\alpha\gamma}^\beta = \pdif{\mbf{g}_\alpha}{y^\gamma}\bcdot \mbf{g}^\beta.
\end{align}
The connection Laplacian acting on a vector is 
\begin{align}
    \Delta_{C} \mbf{v} = \nabla_\alpha \nabla^\alpha v^\beta \mbf{g}_\beta, 
\end{align}
and the Laplace-Beltrami operator acting on a scalar is
\begin{align}
    \Delta_{LB} f = \nabla_\alpha \nabla^\alpha f.  
\end{align}
The deterministic spatial flux can be expressed as
\begin{align}
    \dot{\mbf{x}}_d = v^\alpha \mbf{g}_\alpha,
\end{align}
where
\begin{align}
    v^\alpha  = U p^\alpha + u^\alpha. 
\end{align}

Note that the covariant basis introduced above is not orthonormal in general, so is inconvenient for representing the unit director $\mbf{p}$. 
Another way to formulate differential geometry on a curved surface is to use Cartan's moving frames and differential forms \citep{do2012differential}. 
The basic idea is to introduce a set of orthonormal frames $\{\me_1, \me_2\}$ in the tangent plane and a connection 1-form $\omega_{12}$ (also called the spin-connection in the physics literature, e.g., \cite{kamien2002geometry,turner2010vortices}) which measures the rotation of the frame due to parallel transport on the surface. 
Taking the exterior derivative of $\mbf{p} = \cos\psi \me_1 + \sin\psi \me_2$:
\begin{eqnarray}
    \dd \mbf{p} = -\sin\psi \dd \psi \me_1 + \cos\psi \dd \me_1 + \cos\psi \dd \psi \me_2 + \sin\psi \dd \me_2. 
\end{eqnarray}
In 2D, the only non-zero tangential component in the connection 1-form is
\begin{align}
    \omega_{12} = \dd \me_1 \bcdot \me_2 = -\omega_{21}, 
\end{align}
therefore,
\begin{align}
    \dd \mbf{p} & = -\sin\psi (\dd \psi + \omega_{12}) \me_1 + \cos\psi (\dd \psi + \omega_{12}) \me_2 \\
    & = (\dd \psi + \omega_{12}) \mbf{p}^\perp,
\end{align}
where $\mbf{p}^\perp = -\sin\psi \etheta + \cos\psi \ephi$. 
Pulling back this exterior expression to the deterministic flow yields the relation:
\begin{align}
    \dot{\mbf{p}}_d = [\dot{\psi}_d + \omega_{12}(\dot{\mbf{x}}_d)]\mbf{p}^\perp. 
\end{align}
Equating this expression with the deterministic flux driven by the flow-alignment in the tangent plane:
\begin{align}
    \dot{\mbf{p}}_d = (\mathsfbi{I}_{\,2} - \mbf{p}\mbf{p})\bcdot (\beta\mathsfbi{E}-\mathsfbi{W}) \bcdot\mbf{p},
\end{align}
we obtain the angular flux as
\begin{align}
    \dot{\psi}_d = - \omega_{12}(\dot{\mbf{x}}_d) + \mbf{p}^\perp \bcdot (\beta\mathsfbi{E}-\mathsfbi{W}) \bcdot \mbf{p}. 
\end{align}

In spherical coordinates, we have
\begin{align}
    \mbf{x} = (\sin\theta\cos\phi,\; \sin\theta \sin\phi,\; \cos\theta), 
\end{align}
and 
\begin{align}
    & \mbf{g}_\theta = (\cos\theta\cos\phi,\; \cos\theta\sin\phi,\; -\sin\theta) = \etheta, \\[3pt]
    & \mbf{g}_\phi = (-\sin\theta\sin\phi,\; \sin\theta\cos\phi,\; 0) = \sin\theta\, \ephi,\\
    & \Gamma_{\phi\phi}^\theta = -\sin\theta\cos\theta, \quad \Gamma_{\theta\phi}^\phi = \Gamma_{\phi\theta}^\phi = \cot\theta. 
\end{align}
The connection 1-form is
\begin{align}
    \omega_{12} = \dd\etheta \bcdot \ephi = \cos\theta \,\dd \phi, 
\end{align}
and therefore
\begin{align}
    \dot{\psi}_d = - \cos\theta\, \dot{\phi}_d + \mbf{p}^\perp \bcdot (\beta\mathsfbi{E} - \mathsfbi{W}) \bcdot \mbf{p}. 
\end{align}
The deterministic spatial fluxes can be directly written from the components as
\begin{align}
   & \dot{\theta}_d = U\cos\psi + u^\theta, \\
   & \dot{\phi}_d = U \frac{\sin\psi}{\sin\theta} + u^\phi.
\end{align}

\section{Properties of spin-weighted spherical harmonics}\label{appB}
The properties of SWSHs are summarized in this section, and the reader is referred to other references on the topic for more details \citep{newman1966note,goldberg1967spin,mcewen2011novel,beyer2014numerical,vasil2019tensor,freeden2022spin,price2024differentiable}. 

From the orthonormal frame, $\{\etheta , \ephi \}$ on the tangent bundle to $S^2$, we can introduce a set of null basis vectors: 
\begin{align}
    \mm = \frac{1}{\sqrt{2}}(\etheta + \ii \ephi), \quad \bmm = \frac{1}{\sqrt{2}}(\etheta - \ii \ephi),
\end{align}
satisfying 
\begin{align}
    \mm\bcdot \mm = 0,\quad \mm\bcdot \bmm = 1.
\end{align}
The covariant gradient operator on the unit 2-sphere is
\begin{align}
    \nbp = \etheta \nb_{\etheta} + \ephi \nb_{\ephi},
\end{align}
where
\begin{align}
    & \nb_{\etheta} \etheta = 0, \quad \nb_{\ephi} \etheta = \cot\theta \,\ephi, \\
    & \nb_{\etheta} \ephi = 0, \quad \nb_{\ephi} \ephi = -\cot\theta \,\etheta, 
\end{align}
so 
\begin{align}
    & \nbp \mm = -\frac{\cot\theta}{\sqrt{2}}(\mm - \bmm) \mm, \\
   & \nbp \bmm = \frac{\cot\theta}{\sqrt{2}}(\mm - \bmm) \bmm.
\end{align}
When acting on a function $f_s$ with spin weight $s$:
\begin{align}
    & \mm \bcdot \nbp f_s = - \frac{1}{\sqrt{2}}(\eth f_s - s\cot\theta f_s),\\
    & \bmm \bcdot \nbp f_s = - \frac{1}{\sqrt{2}}(\bar{\eth} f_s + s\cot\theta f_s),
\end{align}
so the covariant gradient of $f_s$ can be represented by
\begin{align}
    \nbp f_s = - \frac{1}{\sqrt{2}}(\eth f_s - s\cot\theta f_s) \bmm 
     - \frac{1}{\sqrt{2}}(\bar{\eth} f_s + s\cot\theta f_s) \mm.
\end{align}
When $f$ is a scalar function with spin weight $s=0$, this gives
\begin{align}
    \nbp f = - \frac{1}{\sqrt{2}}(\eth f \bmm 
     + \bar{\eth} f \mm).
\end{align}
The covariant gradient of a vector field $\mbf{F}$ in the tangent plane can be expressed as
\begin{align}
     \nbp \mbf{F} =& (\nbp F^+) \bmm + F^+ (\nbp \bmm) + (\nbp F^-) \mm + F^- (\nbp \mm) \\
     =& - \frac{1}{\sqrt{2}}(\eth F^+ \bmm \bmm + \eth F^- \bmm \mm + \bar{\eth}F^+ \mm \bmm + \bar{\eth} F^- \mm \mm).
\end{align}

The spin-weighted spherical harmonics are defined as
\begin{align}
    {}_s Y_{lm} = \begin{cases}
    \displaystyle (-1)^s \sqrt{\frac{(l+s)!}{(l-s)!}} \bar{\eth}^{-s} Y_{lm}, \quad -l \leq s \leq 0,\\[2ex]
    \displaystyle \sqrt{\frac{(l-s)!}{(l+s)!}} \eth^s Y_{lm}, \quad 0 \leq s \leq l,
    \end{cases}
\end{align}
where the standard spherical harmonics $Y_{lm}$ follow the convention
\begin{align}
    Y_{lm} = \sqrt{\frac{2l+1}{4\pi} \frac{(l-m)!}{(l+m)!}} P_l^m(\cos\theta) \ee^{\ii m \phi}.
\end{align}
The SWSHs satisfy the conjugate relation 
\begin{align}
    {}_{s}\bar{Y}_{lm}(\theta,\phi) = (-1)^{m+s} {}_{-s}{Y}_{l,-m}(\theta,\phi),
\end{align}
and form a complete, orthonormal basis for spin-weighted functions:
\begin{align}
   & f_s(\theta,\phi) = \sum_{l=|s|}^{\infty} \sum_{m=-l}^l \hat{f}_{lm} \; {}_{s}Y_{lm}(\theta,\phi), \\
   & \hat{f}_{lm} = \int_{S^2} f_s(\theta,\phi) {}_{s}\bar{Y}_{lm}(\theta,\phi) \dd \Omega, 
\end{align}
where $\dd \Omega = \sin\theta \dd\theta \dd \phi$. 
The action of the $\eth,\bar{\eth}$ operators on the SWSHs is given by
\begin{align}
    & \eth {}_sY_{lm} = \sqrt{(l-s)(l+s+1)} {}_{s+1}Y_{lm}, \\
    & \bar{\eth} {}_sY_{lm} = - \sqrt{(l+s)(l-s+1)} {}_{s-1}Y_{lm}.
\end{align}

\section{Derivation of the moment hierarchy equations in terms of SWSHs \label{sec:appendixC}}
Here, we derive the moment hierarchy equation \eqref{eq:Psinequation} from the Fokker-Planck equation \eqref{eq:FPE2}. 
The terms involving self-propulsion read
\begin{align}
     - &\nbp\bcdot (U\mbf{p} \Psi) + \partial_\psi(U\cot\theta \sin\psi \Psi) \\
    &=- U\left[\frac{1}{\sin\theta}\partial_{\theta}(\sin\theta\cos\psi \Psi) + \partial_{\phi} \left(\frac{\sin\psi}{\sin\theta} \Psi\right) \right] + \partial_\psi(U\cot\theta \sin\psi \Psi) \\
    &= \frac{U}{2} \sum_n \left[-(\partial_\theta - \frac{\ii}{\sin\theta} - n \cot\theta) \Psi_n \ee^{\ii (n+1)\psi} - (\partial_\theta + \frac{\ii}{\sin\theta} + n \cot\theta) \Psi_n \ee^{\ii (n-1)\psi} \right] \\
    &= \frac{U}{2} \sum_n (\bar{\eth}\Psi_{n-1} + \eth \Psi_{n+1}) \ee^{\ii n \psi}. 
\end{align}
Similarly, the terms involving advection by the fluid flow are 
\begin{align}
      \!\!- &\nbp\bcdot (\mbf{u} \Psi) + \partial_\psi(\cos\theta u^\phi \Psi) \\
     &\!= - \mbf{u}\bcdot \nbp\Psi + \partial_{\psi}\left[\frac{\cot\theta}{\sqrt{2}\ii}(u^+ - u^-) \Psi \right]  \\
     &\!= \frac{1}{\sqrt{2}}\sum_n \left[-u^+ (\partial_\theta\! -\! \frac{\ii}{\sin\theta} \partial_\phi\! -\! n \cot \theta) \Psi_n \ee^{\ii n \psi} \!-\! u^- (\partial_\theta + \frac{\ii}{\sin\theta} \partial_\phi + n \cot \theta) \Psi_n \ee^{\ii n \psi}\right] \\
     &\!= \frac{1}{\sqrt{2}}\sum_n (u^+ \bar{\eth}\Psi_n + u^- \eth \Psi_n) \ee^{\ii n\psi}. 
\end{align}
The flow-alignment terms are
\begin{align}
    \mbf{p}^\perp \bcdot \mathsfbi{E} \bcdot \mbf{p} = \frac{\ii}{2\sqrt{2}} (\eth u^+ \ee^{-2\ii \psi} - \bar{\eth} u^- \ee^{2\ii\psi})
\end{align}
and
\begin{align}
    \mbf{p}^{\perp} \bcdot \mathsfbi{W} \bcdot \mbf{p} = \frac{\ii}{\sqrt{2}}\eth u^-, 
\end{align}
so the rotational flux term capturing flow alignment is
\begin{align}
    \partial_\psi[\mbf{p}^\perp \bcdot (\beta\mathsfbi{E} - \mathsfbi{W}) \bcdot \mbf{p}] = -\sum_n \frac{n}{2\sqrt{2}}[\beta(\eth u^+ \Psi_{n-2} - \bar{\eth} u^- \Psi_{n+2}) - 2 \eth u^- \Psi_n] \ee^{\ii n \psi}. 
\end{align}
Finally, the translational and rotational diffusion terms are
\begin{align}
    D\Delta_{LB} \Psi = \sum_n \frac{D}{2}(\eth\bar{\eth} + \bar{\eth}\eth) \Psi_n \ee^{\ii n \psi}
\end{align}
and
\begin{align}
    \partial_\psi^2 \Psi = - \sum_n n^2 \Psi_n \ee^{\ii n \psi}. 
\end{align}

\section{Derivation of the force balance equation in terms of SWSHs} \label{appC}
The surface incompressibility condition $\nbp\bcdot\mbf{u}=0$ can be expressed in the null basis as
\begin{align}
    -\frac{1}{\sqrt{2}}(\bar{\eth}u^+ + \eth u^-) = 0.
\end{align}
With the SWSH expansion of $u^\pm$ and the properties of SWSHs, this implies that $u^+_{lm} = u^-_{lm}$. 

Lamb's solution of the Stokes equations \citep{kim2013microhydrodynamics}, keeping only the toroidal part $\sum_l\nb\times(\mbf{x}\xi_l)$, can be projected to the SWSHs as
\begin{align}
    & \mbf{u}^{\text{in}}|_{r=1} = \sum_{l=1}^\infty \sum_{m=-l}^{l} \ii \sqrt{\frac{l(l+1)}{2}} c_{lm}^{\text{in}} ({}_1 Y_{lm}\bmm + {}_{-1} Y_{lm}\mm),\\
    & \mbf{u}^{\text{out}}|_{r=1} = \sum_{l=1}^\infty \sum_{m=-l}^{l} \ii \sqrt{\frac{l(l+1)}{2}} c_{lm}^{\text{out}} ({}_1 Y_{lm}\bmm + {}_{-1} Y_{lm}\mm). 
\end{align}
To satisfy the no-slip boundary condition
\begin{align}
    \mbf{u} = \mbf{u}^{\text{in}}|_{r=1} = \mbf{u}^{\text{out}}|_{r=1},
\end{align}
the coefficients should satisfy
\begin{align}
    c_{lm}^{\text{in}} = c_{lm}^{\text{out}} = -\ii \sqrt{\frac{2}{l(l+1)}} u_{lm}^+. 
\end{align}
Denoting $\mbf{\sigma} = -p\mathsfbi{I}_{\,3} + \lambda (\nb \mbf{u} + \nb\mbf{u}^\top)$ as the bulk stress, the traction forces are given by
\begin{align}
    & \mbf{N}\bcdot \mbf{\sigma}^{\text{in}}|_{r=1} = \lambda^{\text{in}} \sum_{l=1}^\infty \sum_{m=-l}^{l} (l-1) u_{lm}^+ ({}_1 Y_{lm}\bmm + {}_{-1} Y_{lm}\mm), \\
    & \mbf{N}\bcdot \mbf{\sigma}^{\text{out}}|_{r=1} = \lambda^{\text{out}} \sum_{l=1}^\infty \sum_{m=-l}^{l} (-l-2) u_{lm}^+ ({}_1 Y_{lm}\bmm + {}_{-1} Y_{lm}\mm),
\end{align}
and therefore the traction jump is
\begin{align}
    \llbracket\mbf{T}\rrbracket = - \sum_{l=1}^\infty [\lambda^{\text{out}}(l+2) + \lambda^{\text{in}}(l-1)]\sum_{m=-l}^l u_{lm}^+ ({}_1 Y_{lm} \bar{\mbf{m}} + {}_{-1} Y_{lm} \mbf{m}) \eqqcolon T^+ \bmm + T^- \mm.
\end{align}
The surface viscous term reads
\begin{align}
    (\Delta_{C} + K)\mbf{u} = [\frac{1}{2}(\eth\bar{\eth} + \bar{\eth}\eth) + K]u^+ \bmm 
    + [\frac{1}{2}(\eth\bar{\eth} + \bar{\eth}\eth) + K] u^- \mm.
\end{align}
Projecting the force balance on the membrane to the null basis yields:
\begin{align}
    & f^+ + T^+ = - \mathcal{L}u^+ - \frac{1}{\sqrt{2}}\eth \gamma ,\\
    & f^- + T^- = - \mathcal{L}u^- - \frac{1}{\sqrt{2}}\bar{\eth} \gamma, 
\end{align}
where we have defined $\mathcal{L} \coloneqq (\eth\bar{\eth} + \bar{\eth}\eth)/2 + K$. 
Applying $\bar{\eth}$ on both sides of the first equation and $\eth$ on the second equation, and eliminating the surface pressure using $\bar{\eth}\eth \gamma = \eth\bar{\eth}\gamma$, we obtain
\begin{align}
    \bar{\eth}(f^+ + T^+) - \eth(f^- + T^-) = -\bar{\eth}\mathcal{L}u^+ + \eth \mathcal{L} u^-. 
\end{align}
Inserting the SWSH expansions of the active force and traction jump, we arrive at equation \eqref{eq:uPDFcoeff} relating the coefficients $u_{lm}^\pm$ and $a_{lm}^{(\pm 2)}$.

\vspace{0.5cm}

\bibliographystyle{jfm}
\bibliography{jfm}

\end{document}